\documentclass[12pt,preprint]{aastex}

\shorttitle{NLTE UV modeling}
\shortauthors{Short and Hauschildt}

\begin{document}


\title{Non-LTE modeling of the near UV band of late-type stars}


\author{C. Ian Short}
\affil{Department of Astronomy \& Physics and Institute for Computational Astrophysics, Saint Mary's University,
    Halifax, NS, Canada, B3H 3C3}
\email{ishort@ap.smu.ca}

\author{P.H. Hauschildt}
\affil{Hamburger Sternwarte, Gojenbergsweg 112, 21029 Hamburg, Germany}
\email{phauschildt@hs.uni-hamburg.de}




\begin{abstract}

We investigate the ability of both LTE and Non-LTE models to fit the
near UV band absolute flux distribution, $f_\lambda(\lambda)$,
and individual spectral line profiles of three standard 
stars for which high quality spectrophotometry and high resolution
spectroscopy are available: The Sun (G2 V), Arcturus (K2 III), and Procyon (F5 IV-V).  
We investigate 1) the effect of the choice of atomic line list on 
the ability of NLTE models to fit the near UV band $f_\lambda$ level, 2) the 
amount of a hypothesized continuous thermal absorption extinction source required 
to allow NLTE models to fit the observations, and 3) the semi-empirical 
temperature structure, $T_{\rm kin}(\log\tau_{\rm 5000})$, 
required to fit the observations with NLTE models and standard continuous 
near UV extinction.  
We find that all models that are computed with high quality atomic line lists
predict too much
flux in the near UV band for Arcturus, but fit the warmer stars well.
The variance among independent measurements of the
solar irradiance in the near UV is sufficiently large that we cannot
definitely conclude that models predict too much near
UV flux, in contrast to other recent results.
We surmise that the inadequacy of current atmospheric models 
of K giants in the near UV band is best addressed by hypothesizing that there is still 
missing continuous thermal extinction, and that the 
missing near UV extinction becomes more important with decreasing effective 
temperature for spectral classes later than early G, suggesting a molecular origin.

\end{abstract}



\section{Introduction}

 \citet{short96} and \citet{short94} presented an assessment of the ability
of LTE modeling performed with the ATLAS9 code \citet{kurucz92a} to fit the
near UV band of the Sun and the K1.5-2 III giant Arcturus ($\alpha$ Boo, HR5430, 
HD124897), respectively.  The former paper included an assessment of the quality
of various input line lists, and found that the line list that provided the best
match to the detailed high resolution near UV solar spectrum allowed too much
flux to escape in the near UV band, making the model too bright compared to
absolute spectrophotometric measurements in the near UV.  The latter paper
made the case for a hitherto undiscovered source of continuum thermal extinction 
in the near UV band
of early K giants on the grounds that LTE models and synthetic spectra computed
with {\it any} line lists, including the most complete ones, predicted significantly
more flux in the near UV band than was observed.  In the same year as the latter paper, 
\citet{bell_pt94} also examined the 
ability of various line lists to provide a fit to the near UV band solar spectrum,
and concluded that there may still be missing near UV extinction in models
of the solar atmosphere.

 The LTE studies of \citet{short96} and \citet{short94} were improved by 
Non-LTE (NLTE) modeling of Arcturus \citep{shorth03} (hereafter Paper I) and the 
Sun \citep{shorth05} (hereafter Paper II) carried out with the PHOENIX 
multi-purpose atmospheric modeling and spectrum synthesis code \citep{hauschildt_b99}.
The results of Paper I and Paper II were that more realistic NLTE modeling only
made the near UV band discrepancies found with LTE models by \citet{short94} and 
\citet{short96} even worse as a result of NLTE over-ionization of \ion{Fe}{1} 
partially lifting the ``iron curtain'' of near UV \ion{Fe}{1} line extinction, thus
making models even brighter in the near UV.  That  
\ion{Fe}{1} suffers NLTE over-ionization (or, more properly, LTE {\it under}-ionization!)
in late type stars is not a new result (for, example, see \citet{rutten86}), but
Paper I and II presented the first quantitative assessment of the effect on solar 
near UV $f_\lambda$ levels with atomic models of \ion{Fe}{1} and \ion{Fe}{2} containing
over 1000 atomic $E$ levels combined, and accounting for non-local multi-species
coupling due to over-lapping transitions (shared radiation fields) among dozens of
relatively abundant chemical species.

In this paper we investigate two ways of addressing the problem of atmospheric models
of late-type stars being too bright in the near UV band: 1) we revisit the suggestion
of \citet{short94} that there may be missing continuum thermal extinction in the 
near UV, but on the basis of NLTE modeling rather than LTE modeling, and 2) we 
attempt to construct semi-empirical models of the atmospheric temperature
structure $T_{\rm kin}(\log\tau)$ that more closely match the observed near UV 
flux levels.  We also attempt to map out the dependence of a possible source of
missing continuum extinction by investigating its dependence on stellar parameters
among an (admittedly) small set of stars for which there is both high quality 
spectrophotometry that extends into the near UV band, and high quality high
resolution near UV band spectra: namely, the Sun (G2 V), Arcturus (K2 III), and 
Procyon (F5 IV-V, $\alpha$ CMi, HR 2943, HD 61421).     



\section{Near UV band $f_{\lambda}(\lambda)$ distributions}

\subsection{Absolute spectrophotometry}

\subsubsection{The Sun}

We have assembled four relatively recent independent 
measurements of the solar irradiance spectrum in the visible and near UV bands,
 which are presented in Fig. \ref{sun_flxcmpuv}.  The four data sets are most 
discrepant with each other in the near UV band, as described is greater
detail below, and, in the case of the data sets of \citet{necklabs}, 
\citet{burlov95} and \citet{lockwood92}, this may be because the correction
for air mass extinction is more difficult in the near UV than in the visible
band.  The four data gets were re-sampled at the sampling interval of the highest
resolution data set, $\Delta\lambda=3.0$ \AA~ \citep{thuillier03}, and then
each was convolved to the resolution of the most coarsely sampled data-set,
$R\approx70$ at 3500 \AA~ ($\Delta\lambda=50$) \AA~ \citep{burlov95}.  For
this aspect of our investigation, we are concerned with the fit of models
to the overall $f_\lambda$ level on the intermediate-band scale, rather than 
to finer spectral features.

\citet{necklabs} presents a heavily cited construction of the irradiance based on
measurements of the disk center intensity spectrum, $I_\lambda(\mu=1, \lambda)$, 
combined with Fourier transform 
spectrometer (FTS) spectra and center-to-limb variation data obtained
with the McMath solar telescope at Kitt Peak, and is 
the data set to which we compared PHOENIX models in Paper II. 
Their short wavelength data set spans the 3300 to 6300 \AA~ range in 
intervals of 10 \AA.

Like \citet{necklabs}, \citet{burlov95} also constructed 
the irradiance in bands of $\Delta\lambda=50$ \AA~ from 
$I_\lambda(\mu=1, \lambda)$
measurements made in the 3100 to 6850 \AA~ range with the 0.23 m solar 
telescope at Peak Terskol.  They estimate the uncertainty in their measured
$f_\lambda$ values to be $2.5\%$ in the near UV band.
They note that their irradiance values 
are $8\%$ higher in the 3100 to 4000 \AA~ range than those measured by \citet{necklabs} 
(notwithstanding high frequency oscillations in the comparison at the 
wavelengths of the Fraunhofer lines) as seen in their Fig. 11, and 
they conclude that the values of \citet{necklabs} are too low.

\citet{lockwood92} measured the irradiance in the 3295 to 8000 \AA~
range in $\Delta\lambda=4$ \AA~ bands with the 0.6 m Morgan telescope at Lowell 
Observatory using pinhole optics to reproduce a stellar-like image of the Sun, 
and calibrating their measurements against similar measurements of the standard
star Vega.  They estimate the internal error of their $f_\lambda$ values
to be $\approx2\%$.  They also find that their values on wavelength scales
wider than that of Fraunhofer lines in the 3500 to 4500 \AA~
range are as much as $10\%$ larger than those of \citet{necklabs}, and this
discrepancy can clearly be seen in their Fig. 6.

\citet{thuillier03} 
report space based measurements of the near UV and visible band irradiance made 
with the UV and VIS spectrometers
of the SOLSPEC instrument that was part of the ATLAS payloads and the SOSP 
instrument on board the EURECA satellite.  The UV spectra span the 
2000 to 3600 \AA~ range in 4 \AA~ $\Delta\lambda$ intervals, and in the
3400 to 8500 \AA~ range in 10 \AA~ intervals.
Their study includes comparisons between
the SOLSPEC and SOSP instruments and other space-based measurements that
were made at about the same time, and finds agreement among them at the level
of less than $\approx 5\%$.  A special advantage of this data set 
in the near UV
band compared to the others compiled here is that the correction for air mass 
extinction is obviated for space-based observations. 
\citet{thuillier98} reports a comparison between
the visible band data described in \citet{thuillier03} and the data of
\citet{necklabs}, and panel {\it e)} of their Fig. 2 and their Table {\sc V} also 
show that their values are $\approx 4\%$ larger than \citet{necklabs} in the 
3500 to 4000 \AA~ region.

\subsubsection{Arcturus and Procyon}

\citet{burn} presents a compilation of $f_\lambda$ distributions of a large 
sample of stars, including
1) their own measurements made at the Crimean Astrophysical Observatory,  
2) measurements from the ``Sternberg spectrophotometric catalogue'' 
(see \citet{glushneva84} and references therein) consisting of data taken 
from 1970 to 1984 with the
40 cm and 60 cm telescopes of the Crimean Station of the Sternberg Astronomical
 Institute and the 50 cm telescope of the Fessenkov Astrophysical Institute, and 
3) measurements of \citet{kharitonov78} taken with the 50 cm Cassegrain telescope
and spectrum scanner with a photo-multiplier tube taken at the Fessenkov 
Astrophysical Institute from 1968 to 1986.
These data sets
all generally cover the $\lambda$ range 3200 to 8000 \AA~ with 
$\Delta\lambda=25$ \AA~ or $50$ \AA, and have been re-reduced to a common
spectrophotometric system (the ``Chilean system'') by \citet{burn}. 
\citet{glushneva84} and \citet{kharitonov78} state that the internal
accuracy of their $f_\lambda$ measurements is $\approx3.5\%$.  Arcturus and Procyon
are among the stars for which data was reported in all three of these sources.

Additionally, for Procyon, which is the brightest of the three objects in the UV band, 
we were also able to obtain a mid-UV spectrum recorded with the International
Ultraviolet Explorer (IUE) space based observatory in the ``long wavelength'' band
(LWR09108 in the NEWSIPS IUE archive).
This data set covers the $\lambda$ range from 1850 \AA~ to 3350 \AA, and provides
a check on the ability of the models to fit the $f_\lambda$ distribution at shorter 
wavelengths.

\subsection{High resolution spectroscopy}

\paragraph{Sun }
\citet{kurucz_fbt} obtained a spectral flux atlas of the Sun with the Fourier Transform 
Spectrometer at the McMath/Pierce Solar Telescope at Kitt Peak National
Observatory (KPNO) in the $\lambda$ range of 2960 to 13000 \AA.  The spectral
resolution, $R$, is $\approx 800\, 000$ in the near UV, but we have degraded
the data set to $R=200\,000$ by convolution with a Gaussian kernel to reproduce
a more typical ``high resolution'' stellar spectrum, and to aid comparison with
synthetic spectra computed with a realistic sampling, $\Delta\lambda$.  The
spectrum is useful for the present investigation as long
as $R$ is sufficiently large for weak spectral lines to be fully resolved.
\citet{wallace} independently obtained a full disk spectrum of the Sun with
the same instrument and similar $R$ value.  We also include the latter
in our comparison to assess the variance among data sets.  We note that the
latter spectrum is darker by about 0.02 to 0.05$f_\lambda^{\rm c}$ (where  
$f_\lambda^{\rm c}$ is the continuum flux level) than the atlas of
\citet{kurucz_fbt} in the cores of some, but not all, strong features such as 
the CN band head
at $\lambda3883$ (see Fig. \ref{sun_atlas1}) and the \ion{Ca}{2} H and K lines,
although the two spectra are in close agreement for weaker features, and in the
pseudo-continuum.  This non-systematic variance complicates the assessment of 
models based on high resolution spectra because it is precisely in the
cores of strong features that the role of the continuous extinction
manifests itself.

\paragraph{Arcturus }

\citet{hinkle_wvh00} have obtained a high resolution ($R=150\, 000$) spectrum
with the KPNO Coud\'e feed telescope and echelle 
spectrograph.  We note that the observed spectrum only covers wavelengths down to 
$\approx3730$ \AA.
Because \citet{short96} concluded that a surprisingly large amount
of extra near UV opacity was needed to fit the high resolution photographic 
spectrum of \citet{griffin68},
we also incorporate a comparison of the observed spectrum of \citet{griffin68}
(also of $R\approx 150\, 000$ in the near UV) to that of
\citep{hinkle_wvh00}.  We find that the two observed spectra are in close agreement,
except that in the spectrum of \citet{griffin68} strong absorption features tend to be 
not as deep as those of the spectrum of \citet{hinkle_wvh00}.  It should be noted that
\citet{griffin68} commented on the difficulty and subjectivity of establishing
a continuum rectification ($f_\lambda^{\rm c}$ level) in the blue and near UV 
band as a result of over-blanketing
by spectral lines, particularly in the vicinity of the \ion{Ca}{2} H and K lines
($\lambda 3934$ and  $\lambda 3968$) as a result of the broad Van der Waals damping
wings that affect $f_\lambda$ throughout the 3900 to 4000 \AA~ range.  Given the 
difficulties of calibrating photographic
data, and correcting it for background scattered light, as well as the uncertainty in
continuum placement, the discrepancy is not surprising.
However, given that \citet{short96} based their conclusions on their result that
 synthetic spectra 
have spectral absorption features that were too strong, it behooves us to investigate
the extent to which that conclusion may be weakened by comparison to the \citep{hinkle_wvh00}
spectrum.

\paragraph{Procyon }
 \citet{griffinproc} have created a high resolution spectral atlas of Procyon
in the $\lambda$ 3150 to 7500 \AA~ range based on spectra obtained with the
Coud\'e spectrograph of the Mount Wilson 100-inch reflector.  We note that,
like the spectral atlas of \citet{griffin68}, the detector was a photographic
emulsion.  Based on the instrumental profile described in \citet{maeckle_gh75},
we estimate that $R \approx 75\, 000$ in the near UV band (FWHM $\approx 0.05 \AA$). 
The same difficulty of continuum rectification in the near UV band that was
discussed above for the Arcturus spectroscopy also applies to Procyon, as noted
in \citet{griffinproc}.

\section{Models}

\subsection{Modeling}

We have computed model structures and synthetic spectra for the three
program stars that explore a number of variations in the modeling.  
The details of the modeling are as described in Papers I and II.  
The same modeling procedure is used for all three objects, except that
Arcturus is modeled with spherical geometry because of its lower surface 
gravity ($\log g \le 2$), whereas the Sun and Procyon are modeled with
plane-parallel geometry.  
We note that our models are in radiative-convective equilibrium (RE), and, thus,
do not exhibit chromospheric or coronal heating in their outer atmospheres,
which is caused by non-radiative processes.  Therefore, our models do
not have inversions in the $T_{\rm kin}(\log\tau)$ structure, and cannot
reproduce near UV spectral features that are chromospheric in origin (eg. 
the \ion{Ca}{2} H \& K line cores).  However, for $\lambda > 3000$ \AA, the
flux in the continuum, and in weak to moderately strong spectral lines,
arises from layers of the atmosphere that are well below the depth of
temperature minimum, $\log\tau_{\rm 5000}(T_{\rm min})$, in semi-empirical
chromospheric models of the Sun (see \citet{fontenla} and references therein)
 and Arcturus (see \citet{ayres} and references therein).  Therefore,
as long as we avoid any attempt to interpret the fit to the cores of
highly opaque lines where the flux arises from the outer atmosphere, then
the $T_{\rm kin}(\log\tau)$ structure of our models should be relevant to
the spectral features being studied.  Similarly, our models are static and 
horizontally homogeneous, and so do not account for the complex spatial
and temporal structures on a variety of scales that have been observed in
the solar atmosphere.  With the exception of sunspots and the granulation pattern,
these structures (plages, faculae, \ion{Ca}{2} K bright points) exist
in the chromospheric layers and are probably related to the non-radiative 
chromospheric heating mechanism, and, as mentioned above, our models do
not account for the chromospheric $T_{\rm kin}(\log\tau)$ structure in any 
case.

\subsection{Stellar parameters}

In Table \ref{stelpar} we present the values of the stellar atmospheric
parameters used in our modeling.  
Our parameters for Arcturus and the Sun are those of Papers I and II, 
respectively, and justification for our choices can be found there.
We re-iterate here that we have chosen to investigate solar models
that are based on the ``traditional'' abundances, $[A/H]$, of \citet{grev_ns92} rather 
than the newer abundances of \citet{asplund00} derived from horizontally
inhomogeneous models for consistency with the horizontally homogeneous 
solar atmospheric modeling of \citet{kurucz92a}.
\citet{jaufdenberg} recently carried out a simultaneous atmospheric model fit
 to the spectrophotometric $f_\lambda$ distributions and near IR band
interferometry for Procyon, and included multi-component modeling with
combinations of synthetic $f_\lambda$ distributions from 1D atmospheric
models computed with PHOENIX to approximately simulate the effect of 
horizontal inhomogeneity.
They find best fit values for the stellar parameters of $T_{\rm eff}$ in
the range $6510$ to $6515$ K, $\log g$ equal to $3.95 \pm 0.02$, 
metallic $[{{\rm A}\over{\rm H}}]$
 equal to 0.0, a depth-independent value of the microturbulent velocity 
dispersion, $X_{\rm T}$, equal to $2$ km s$^{\rm -1}$, and a value of
the mixing length used in the approximate treatment of convection,
$l/H_{\rm P}$, in the range 0.5 to 1.25, where $H_{\rm P}$ is the pressure
scale height.  (They also find a value of the radius,
$R$, equal to $1.46\times 10^{11}$ cm.)

\subsection{Spectrum synthesis}

In every case the modeling is self-consistent in that the atmospheric structure 
was re-converged before computing the synthetic $f_\lambda$ distribution
for that model.  For comparison to high resolution spectroscopy, we also
produced continuum rectified synthetic spectra by dividing the synthetic
$f_\lambda$ distribution by a corresponding synthetic pure continuum
distribution computed without line extinction.  The synthetic spectra 
were computed with a sampling, $\Delta\lambda$, ranging from 0.015 \AA~
at $\lambda=3000$ \AA~ to 0.04 \AA~ at $\lambda=8000$ \AA~, corresponding
to a spectral resolution, $R$, of $\approx 250\,000$.  This ensured
that weak spectral lines were at least critically sampled.  
The spectra were then resampled and 
convolved with an appropriate kernel to match the $R$ value of the 
observed spectra to which they were being compared.  

\subsection{Models}

Model LTE-big is computed
in LTE using the ``big'' input line list that includes the full set of
spectral absorption lines described by \citet{kurucz92a}, including millions
of atomic lines, largely of Fe-group elements, that were theoretically predicted, 
but have not been observed in the laboratory.  \citet{kurucz92a} found that this
big line list was necessary to provide sufficient line blanketing of the
synthetic $f_\lambda$ distribution in the near UV band of 
a solar atmospheric model to provide a good match to the observed $f_\lambda$ 
distribution of \citet{necklabs}.  This is the line list that was used in 
the LTE modeling presented in Paper II.

Model LTE-small is the same as LTE-big except that the input line list is the 
``small'' list that was used in the ATLAS models of Kurucz prior to the expansion
of \citet{kurucz92a} described above (see, for example, \citet{kuruczp75}).  
This is the line list that was used in 
the NLTE modeling presented in Paper II, and is the only line 
list that is consistent with the NLTE models produced with PHOENIX.

Model NLTE is the same as LTE-small except that the coupled NLTE statistical 
equilibrium rate equations and radiative transfer equation are solved
self-consistently for the lowest one to three ionization stages of nineteen
astrophysically important light metals up to and including the Fe-group 
elements.
This is the same model as the one designated NLTE$_{\rm Fe}$ in Paper II,
and the particular species included in the NLTE treatment, and the
details of the atomic models used, are those
described in Table 1 of that paper.

Model NLTE-extra is the same as model NLTE except that the continuous
mass extinction co-efficient, $\kappa_\lambda^{\rm c}$, has been 
enhanced {\it ad hoc} by a factor that differs for each star, as described in 
Section \ref{sfudge}.

Model LTE-small-extra is the same as model LTE-small, with the addition of the same
amount of extra extinction that was added to the NLTE-extra model.  Model LTE-small-extra
allows an assessment of NLTE effects in the presence of extra near UV extinction by 
comparison to the NLTE-extra model
with all other variables the same, including the UV $\kappa_\lambda^{\rm c}$ sources.

Model NLTE-cool is the same as model NLTE except that the kinetic temperature
structure, $T_{\rm kin}(\tau_{\rm 5000})$, has been adjusted, {\it ad hoc},
as described in Section \ref{scool}.  The Sun was the only object for which 
this variation was either feasible or necessary.

\subsection{{\it Ad hoc} extinction enhancement \label{sfudge}}

For each of the program stars, we produce NLTE-extra models
by enhancing, {\it ad hoc}, the value of the continuous
mass extinction co-efficient, $\kappa_\lambda^{\rm c}$, in the
wavelength range where the discrepancy with the observed $f_\lambda$ 
distribution was most severe, which was $3000 - 4200$ \AA~ in all cases.  
The $\kappa_\lambda^{\rm c}(\lambda)$ distribution was enhanced in 
the affected $\Delta\lambda$ range by adding to $\kappa_\lambda^{\rm c}$
the value of $f\times\kappa_\lambda^{\rm c, T}$, where 
$f\times\kappa_\lambda^{\rm c, T}$ is the continuum
thermal absorption coefficient computed in LTE, and $f$ is a 
$\lambda$-independent {\it ad hoc} factor.  
The $\kappa_\lambda^{\rm c, T}$ expresses the continuum extinction 
caused by the thermal destruction of photons only, as opposed to scattering
extinction; {\it ie.} 
$\kappa_\lambda^{\rm c} = \kappa_\lambda^{\rm c, T} + \sigma_\lambda^{\rm c}$, 
where, $\sigma_\lambda^{\rm c}$ is the continuum
scattering extinction coefficient.
As a result, this enhancement models the effect of an as yet unaccounted
for source of extinction that is purely thermal in that it only destroys
and thermally emits photons rather than scattering them.  The value of 
$f$ was chosen separately for each program star based on visual
inspection of the comparison of the observed and synthetic NLTE $f_\lambda$ 
distributions, and is reported in Table \ref{tabfudge}.  We note that
given the speculative nature of this study, it was not our intention 
to determine the precise value of $f$ that minimizes a statistical 
measure of the goodness of the $f_\lambda$ fit, but, rather, to investigate
whether such an {\it ad hoc} $\kappa_\lambda^{\rm c}$ enhancement
can plausibly improve the fit of NLTE models to both the absolute
$f_\lambda$ distribution and the rectified high resolution spectrum.  


The models with enhanced $\kappa_\lambda^{\rm c}$ were then re-converged
before computing the synthetic $f_\lambda$ distribution so that the models
are internally consistent.  We find that the
thermal equilibrium $T_{\rm kin}(\tau)$ structure of both the Sun and Arcturus,
for both LTE and NLTE models, is negligibly affected by
$\kappa_\lambda^{\rm c}$ enhancement below $\lambda$ values of 4200\AA~
because the fraction of the star's bolometric flux being transferred in the UV 
band is small.  \citet{short94} found the same result for Arcturus with
LTE ATLAS9 modeling.

\subsection{{\it Ad hoc} $T_{\rm kin}$ adjustment \label{scool}}

  We produce the NLTE-cool model by adjusting the $T_{\rm kin}(\log\tau_{\rm 5000})$
structure of the NLTE model downward in value by an offset, 
$\Delta T_{\rm kin}(\log\tau_{\rm 5000})$, that has its maximum amplitude, 
$\Delta T_{\rm kin}(-\infty)$, at
the top of the atmospheric model, which effectively has $\log\tau_{\rm 5000}=-6$, 
and decreases linearly
in value with increasing standard continuum logarithmic optical depth, $\log\tau_{\rm 5000}$, to 0 
near standard optical depth unity ($\log\tau_{\rm 5000}=0$), from around which depth the 
emergent visible and near IR band continuum flux ($f_\lambda^{\rm c}$) arises.  The
temperature offset at each depth in the model is computed as

$\Delta T_{\rm kin}(\log\tau_{\rm 5000})=m\times (\log\tau_{\rm 5000} - (-6)) + \Delta T_{\rm kin}(-\infty)$,

restricted to the range $-6<\log\tau_{\rm 5000}<0$, where the slope, $m$, is given by 
$\Delta T_{\rm kin}(-\infty)/(-6)$, and $\Delta T_{\rm kin}(-\infty) < 0$ because
we are always cooling the atmosphere rather than heating it.   
The $\log\tau_{\rm 5000} > 0$ depth range is where alterations to the 
$T_{\rm kin}(\log\tau_{\rm 5000})$ structure would most clearly amount to changing the 
{\it de facto} $T_{\rm eff}$ value of the model as determined by the measured continuum
flux distribution, $f_{\lambda}^{\rm c}$.  Therefore, to maintain some distinction
between perturbations of $T_{\rm kin}(\log\tau_{\rm 5000})$ for a model of given $T_{\rm eff}$
and perturbations to the $T_{\rm eff}$ value itself, we restrict $T_{\rm kin}$ alterations to 
the $\log\tau_{\rm 5000} < 0$ range. 
The coupled equation of state (EOS), hydrostatic equilibrium (HSE), and the NLTE statistical equilibrium 
rate 
equations, and radiative transfer equation, were reconverged for the resulting
model, but with the temperature correction procedure disabled so that a 
self-consistent model
with the new $T_{\rm kin}(\log\tau)$ structure was computed, then a synthetic
$f_\lambda$ distribution for this model was computed.  
The value of $\Delta T_{\rm kin}(-\infty)$ was chosen by visual inspection of the 
fit of the resulting synthetic $f_\lambda$ distribution to the observed near UV band
$f_\lambda$ values.  
This procedure is effectively semi-empirical modeling of the 
atmospheric $T_{\rm kin}$ structure with the near UV band $f_\lambda$ distribution as the sole 
diagnostic.  
For the Sun, $\Delta T_{kin}(-\infty)=-400$ K.  For Arcturus, the most extreme
value of $\Delta T_{kin}(-\infty)$ that we were able to investigate was 1500 K,
and we found that there 
was no value of $\Delta T_{kin}(-\infty)$ that would sufficiently reduce
the computed near UV $f_\lambda$ distribution without the $T_{\rm kin}(\log\tau)$ structure
becoming so extreme that the fit to the well observed visible band $f_\lambda$ distribution was 
lost, and, for the most extreme values, the population re-convergence became numerically unstable.

\section{The Sun}

\subsection{Spectrophotometric $f_\lambda$ distribution \label{sunabsolut}}

\begin{figure}
\plotone{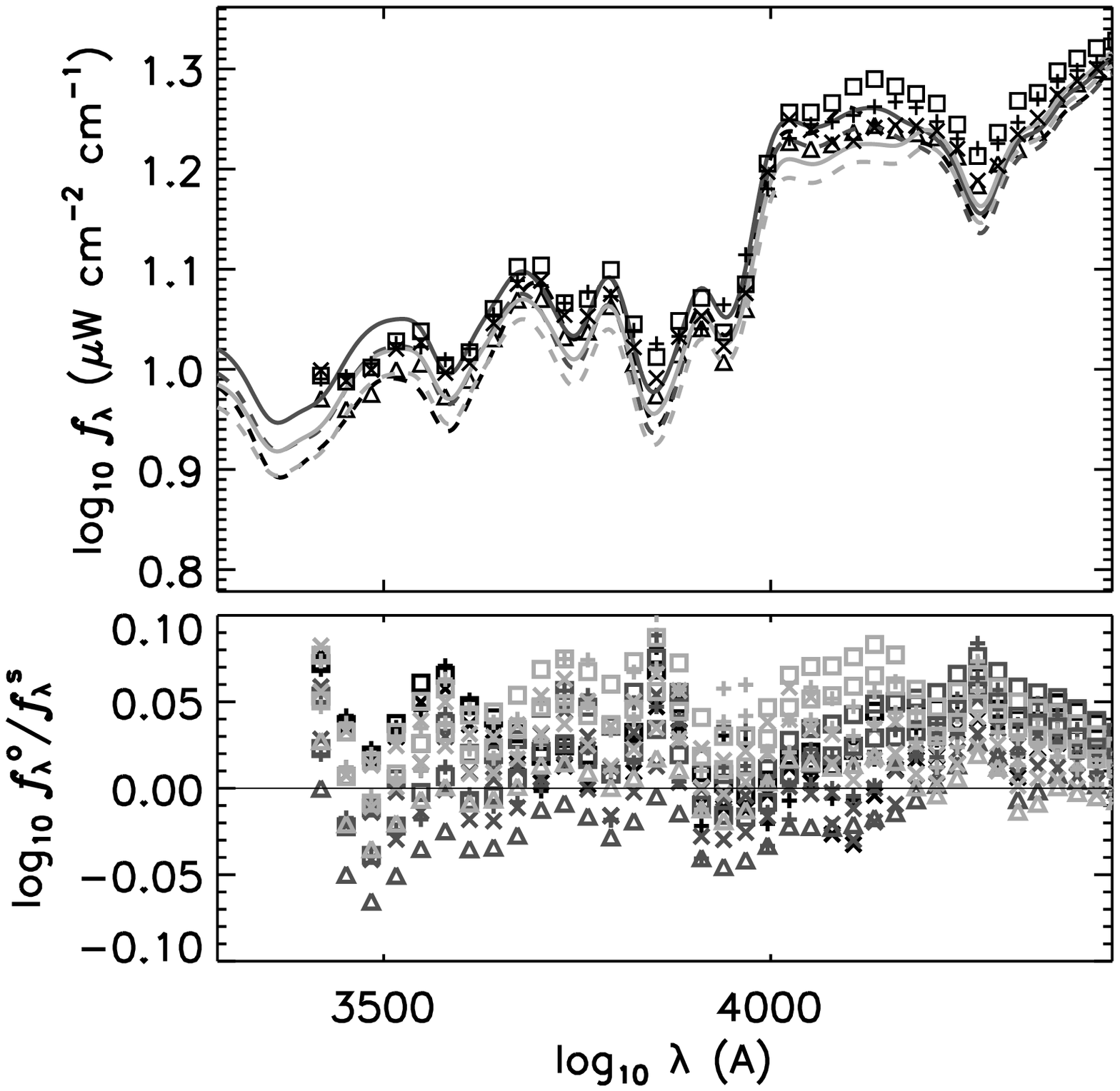}
\caption{Sun: Comparison of observed and computed near 
UV band logarithmic flux
distributions ($\log f_\lambda(\log\lambda)$).  
Upper panel: Direct comparison.  
Lower panel: Difference between
observed ($f_\lambda^{\rm o}$) and synthetic ($f_\lambda^{\rm s}$) spectra 
($\log f_\lambda^{\rm o}/f_\lambda^{\rm s} = \log (f_\lambda^{\rm o} - f_\lambda^{\rm s}$)). 
Observed spectra: \citet{necklabs}: triangles; \citet{burlov95}: crosses; 
\citet{lockwood92}: squares; \citet{thuillier03}: Xs.  Computed spectra:
LTE: dashed lines; NLTE: solid lines.
LTE-big model (same as LTE 
synthetic spectrum of Paper II \citep{shorth05}):
black dashed line; LTE-small model: dark gray dashed line; LTE-small-extra model:
light gray dashed line; NLTE model: dark gray solid line; NLTE-extra model: light gray
solid line.  Note that, for simplicity, we only include 
$\log f_\lambda^{\rm o}/f_\lambda^{\rm s}$ for the NLTE and NLTE-extra models 
in the lower panel.
\label{sun_flxcmpuv}}
\end{figure}

Fig. \ref{sun_flxcmpuv} shows the comparison of all four observed
solar $f_\lambda$ distributions with with the computed $f_\lambda$
distributions from models with various treatments of the thermodynamic
equilibrium and $\kappa_\lambda^{\rm c}$.  Table \ref{tabsunstats}
presents, as a quantitative measure of the relative goodness of fit, the 
root mean square (RMS) of the deviation of the 
synthetic from the observed $f_\lambda$ distribution relative to
the observed distribution, $\sigma$, computed from interpolated
pixel-wise relative deviations in the 3500 to 4000 \AA~ range,

$\sigma^2=\sum_{\rm i}^{\rm N} {\{(f_\lambda^{\rm s}-f_\lambda^{\rm o})/f_\lambda^{\rm o}}\}^2/N$,

where $f_\lambda^{\rm s}$ and $f_\lambda^{\rm o}$ are the synthetic 
and observed $f_\lambda$ distributions, respectively, and $N$ is the 
number of elements of the common $\lambda$ sampling to which all
$f_\lambda$ distributions have been interpolated.  The $\sigma$ values
have been quoted only to the first distinguishing digit.

As reported in Paper II, model LTE-big
reproduces the observed $f_\lambda$ relatively well, thus confirming
the results of the ATLAS modeling of \citet{kurucz92a}.  Also, as 
reported in Paper II, the NLTE model significantly over-predicts 
the $f_\lambda$ values in the near UV band.  In Paper II we
attributed the entire discrepancy between the $f_\lambda$ values of
the LTE-big and NLTE models to NLTE effects, particularly NLTE over-ionization
of \ion{Fe}{1} by the UV band radiation field.  However, a
factor that was neglected in Paper II is that their LTE model
incorporated the ``big'' atomic line list of \citet{kurucz92a} ({\it ie.}
their LTE model is the same as our LTE-big model), whereas
their NLTE model only incorporates the ``small'' line list that pre-dated
the expansion of \citet{kurucz92a}.  Therefore, some of the difference
between the LTE and NLTE$_{\rm Fe}$ $f_\lambda$ distributions is caused
by the much larger number of lines included in the LTE model compared to 
the NLTE model.

\begin{figure}
\plotone{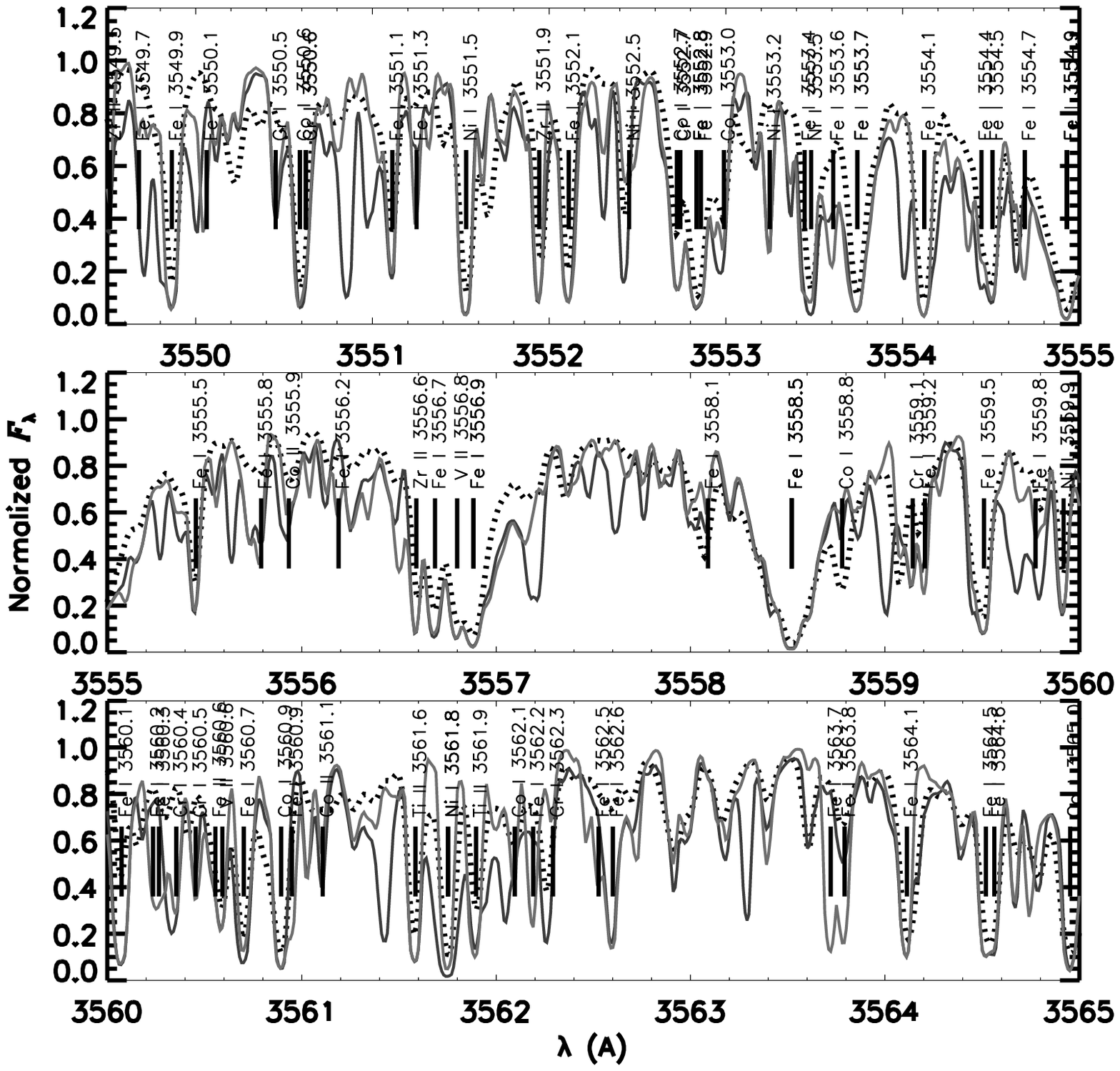}
\caption{Sun: Comparison of observed spectrum of \citet{wallace} (dotted line) 
with computed LTE high resolution flux
spectra ($f_\lambda(\lambda)$) with various atomic line lists.  (The 
observed spectrum of \citet{kurucz_fbt} does not extend down to
this $\lambda$ range.)  Model LTE-big: dark line; model LTE-small: light line.  
Note the relative quality of the model fits around \ion{Fe}{1} $\lambda 3549.7$, 
the feature around $\lambda 3550.8$, around 
\ion{Ni}{1} $\lambda 3552.5$, in the far red wing of the 
\ion{Fe}{1} $\lambda 3554.9$ line, and the features (unidentified in the
``big'' line list) around $\lambda$ 3557.2, 3557.7, 3559, 3561.5, and 3563.3, 
where the LTE-big model predicts
spectral line absorption that is much stronger than that of either
the observed spectrum, or the synthetic spectrum of the LTE-small model.  
The feature around $\lambda 3550.8$ is not identified in the ``big'' line list.
\label{atlasbig1}}
\end{figure}

The model atoms used in the calculation of the NLTE line 
extinction in PHOENIX must be consistent with the line list because 
any transitions not included in the NLTE statistical equilibrium rate 
equations are taken from the line list.  Therefore, it is not currently 
possible to compute a NLTE model with the ``big'' line list with PHOENIX.
To remove the choice of line list as a variable, we have 
computed our LTE-small model, which is more directly comparable to our NLTE
model, and to the NLTE$_{\rm Fe}$ model of Paper II, than the LTE
model of Paper II is.  In addition to the requirement of consistency,
there is some justification for investigating the fit of models computed
with the ``small'' line list in that the ``big'' line list has been controversial.
The ``big'' list includes millions of transitions that were theoretically
predicted using an approximate treatment of the quantum mechanical atomic
physics, and, although these lines improved the broad-band fit to the solar
UV $f_\lambda$ distribution, comparison to high resolution solar
UV spectra indicate that many of the $gf$ values and transition energies
are inaccurate, and that, moreover, many of the lines in the ``big'' line
list seem to not actually exist \citep{bell_pt94}.  This very important
point is also demonstrated again here in Fig. \ref{atlasbig1}.  In general,
 discrepancies between observed and synthetic spectra may be caused by 
errors in the $\lambda$ values in the line list for individual lines that really 
exist.  However, we note
that the careful analysis by \citet{bell_pt94} lead them to conclude that
the ``big'' list really does {\it systematically} contain many lines of
detectable strength that are
either unobserved, or are much weaker, in the observed solar near UV  and blue
spectrum, whereas the reverse is not true, and this is also consistent with our
experience.  
Fig. \ref{atlasbig1} shows the comparison of the observed 
rectified high resolution solar spectrum of \citet{wallace} (the
spectrum of \citet{kurucz_fbt} does not extend down into this $\lambda$
range) and synthetic
spectra computed with the models LTE-big and LTE-small in the $\lambda 3540$ to
$3570$ region.  Examples of where the ``big'' line list leads to synthetic
spectra with lines that are either too strong, or are altogether unobserved,
are scattered through the near UV band, but this spectral range illustrates 
a number of easily discernible examples well.  This point has been made
by both \citet{bell_pt94} and \citet{short96}, but is important enough
to be worth demonstrating here.  On the this basis we proceed on the assumption 
that the ``big'' line list
is less realistic than the ``small'' line list.

\paragraph{}

The LTE-small synthetic $f_\lambda$ distribution
is also shown in Fig. \ref{sun_flxcmpuv}, where it can be seen that it, like
the NLTE synthetic $f_\lambda$ distribution, also over-predicts the
observed flux, although by not as much.  Therefore, the near UV band discrepancy  
between LTE and NLTE models reported in Paper II 
is reduced when consistent line lists are used.  We conclude that the
effect of NLTE \ion{Fe}{1} over-ionization, while still significant,
is not as large as that reported in Paper II. 

\paragraph{}

In Paper II we concluded on the basis of comparing synthetic $f_\lambda$ distributions
to that of \citet{necklabs} that NLTE solar atmospheric models over-predict the
near UV band flux by about $10\%$.
However, including other solar irradiance measurements in the comparison weakens
this conclusion.  Fig. \ref{sun_flxcmpuv} shows that the NLTE synthetic $f_\lambda$
distribution lies at the upper end, but within the range of, the $f_\lambda$ values 
spanned by the four observed distributions, and 
is generally consistent with that measured by \citet{lockwood92}, except for 
a region around 3500 \AA.  We also note that the LTE-big $f_\lambda$ distribution,
although consistent with the data of \citet{necklabs},
is found to lie at the bottom of the range spanned by the four data sets,
suggesting that the LTE-big model actually under-predicts the near UV band flux,
in contradiction to the conclusions drawn by both \citet{kurucz92a} and in Paper II.
By contrast, the LTE-small model, being less bright in the UV band than the NLTE model,
lies well within the range spanned by the four data sets.
This suggests that the ``big'' line list of \citet{kurucz92a} actually contains too
much line opacity to reproduce the observed near UV $f_\lambda$ level, which is 
consistent with the conclusions drawn by \citet{bell_pt94} based in high resolution
spectroscopy.  As a result, given the range spanned by the four data sets, there is
no observational basis for claiming that the, presumably more realistic, NLTE model fits  
better than the LTE-small model, and there is marginal evidence that it provide a {\it worse}
fit.  This speaks to the need for a more precise observational determination of the
UV band $f_\lambda$ values.  Interestingly, the conclusion of both \citet{kurucz92a} and 
Paper II that the LTE-big model provides the best fit may be due to an unfortunate
conspiracy of errors: the observed near UV $f_\lambda$ values of \citet{necklabs} are 
too low, if one were to assume that the other data sets are more accurate, and, at the same time, the ``big''
line list contains spurious extra near UV line opacity that produces an LTE synthetic $f_\lambda$
distribution that is also too low, thus apparently providing a good ``fit'' to 
the data that under-estimate the true flux level.

\paragraph{}

As can be seen from Table \ref{tabsunstats}, the NLTE-extra synthetic 
$f_\lambda$ distribution, by design, provides closer agreement to the dataset 
of \citet{necklabs}, whereas the NLTE model with no additional extinction
provides closer agreement to the data sets of \citet{burlov95} and 
\citet{lockwood92}.  The data set of \citet{thuillier03} does not discriminate 
clearly between the two models.
This may be interpreted as marginal evidence for missing 
near UV band continuous thermal absorption opacity with an enhancement factor, $f$,
of approximately 0.15, but, that conclusion depends on which of the observed $f_\lambda$
distributions one uses in the comparison.  We note that the NLTE-extra 
$f_\lambda$ distribution retains the good
agreement with observed $f_\lambda$ values at longer wavelengths in the visible band.
This is because the fraction of Sun's bolometric flux passing though the $3000-4200$ \AA~
region is small enough that the $\kappa_\lambda^{\rm c}$ enhancement does not affect
the model $T_{\rm kin}(\tau)$ structure.    

\paragraph{}

For $3000 < \lambda < 3600$ \AA, 
the LTE-small-extra spectrum tracks the LTE-big spectrum
closely.  This is not surprising given that the value of $f$ was 
tuned to compensate for the lack of line opacity in the "big" line list so as to 
restore the good fit to $f_\lambda$ provided by the latter.  However, we note
that in the region of $3700 < \lambda > 4200$ \AA~ the LTE-small-extra spectrum 
falls significantly below that of the LTE-big model, indicating that the effect of
extra $\kappa_\lambda^{\rm c}$ does not exactly mimic that of the additional line opacity
in the ``big'' list.  However, we note that we arbitrarily and somewhat crudely
made $f$ independent of $\lambda$ in the 3000 to 4200 \AA~ region, 
so we should not expect 
the LTE-small-extra model to reproduce the spectrum of the LTE-big model
throughout that entire range.  

\paragraph{}

The existence of additional near UV continuous extinction, if real, could, in principle, affect 
NLTE models by extinguishing the near UV radiation field that plays such an
important role in, among other things, the well-known NLTE over-photo-ionization of 
\ion{Fe}{1} (see, for example, Paper II, \citet{rutten86}). 
An assessment of the impact of $f\times\kappa_\lambda^{\rm c, T}$ on NLTE effects can be
made by comparing the synthetic spectra computed with the LTE-small-extra and NLTE-extra
models in Fig. \ref{sun_flxcmpuv}.  We note that,
to a first approximation, the difference between the spectra of the NLTE and
LTE-small models, and that between the NLTE-extra and LTE-small-extra models,
is about the same throughout the 3000 to 4200 \AA~ range.  We conclude that 
the magnitude of $f\times\kappa_\lambda^{\rm c, T}$ is not great enough to significantly
alter the character of the NLTE effects on the solar $f_\lambda$ distribution.
This may not be surprising given the modest value of $f$ (0.15).  
A more careful and detailed analysis of the impact of $f\times\kappa_\lambda^{\rm c, T}$ 
on the NLTE equilibrium of the solar atmosphere and on the detailed solar
spectrum is beyond the scope of 
this investigation, but may be warranted if evidence mounts that the 
$f\times\kappa_\lambda^{\rm c}$ source really exists. 

\paragraph{}

\begin{figure}
\plotone{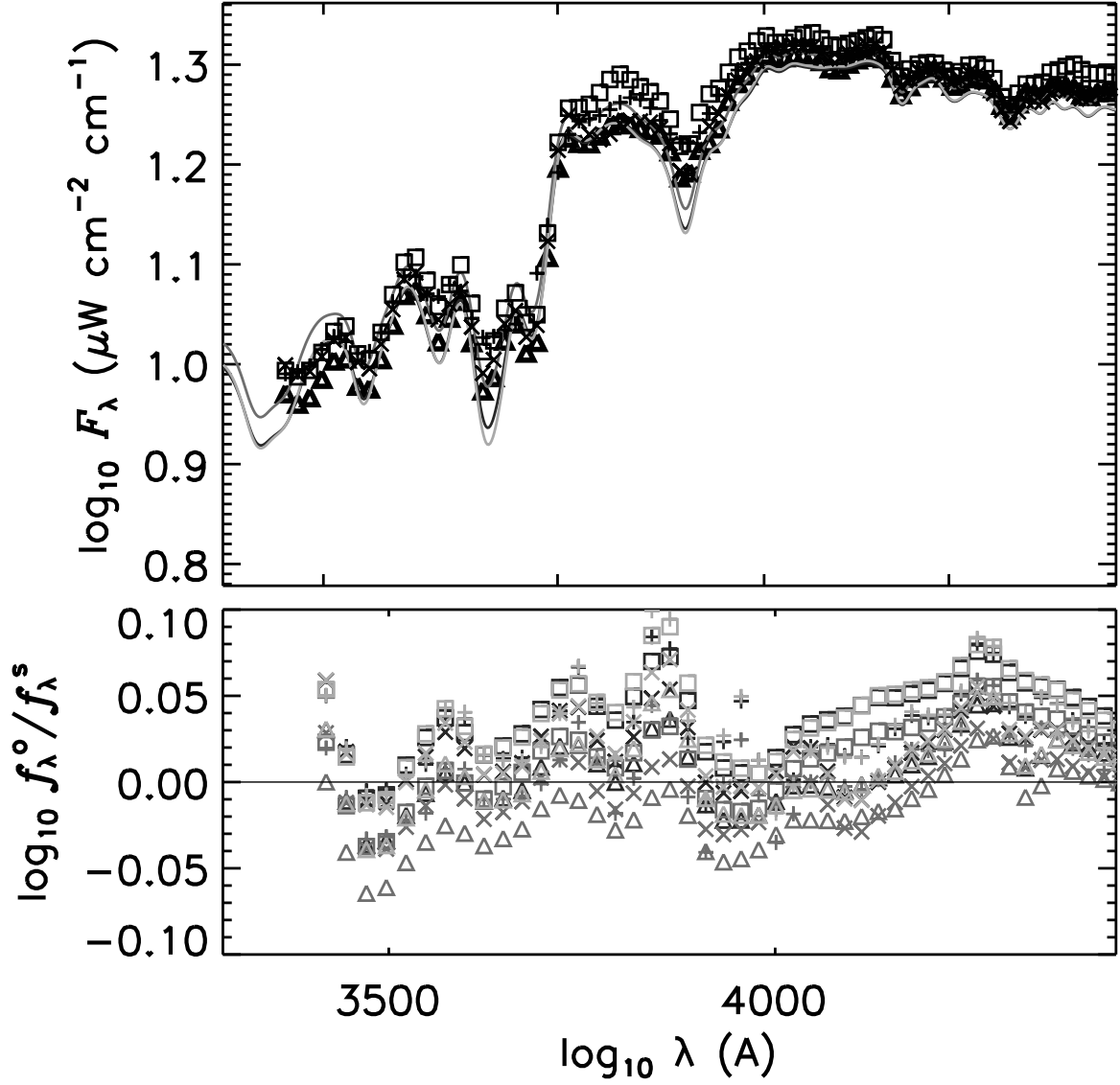}
\caption{Same as Fig. \ref{sun_flxcmpuv}, except that the models explore
the variation in $T_{\rm kin}(\tau_{\rm 5000})$ structure.  LTE-small: 
darkest color; NLTE model: medium gray color; NLTE-cool model: lightest color. 
\label{sun_flxcmpuv2}}
\end{figure}

Fig. \ref{sun_flxcmpuv2} shows the comparison between observed and synthetic
$f_\lambda$ distributions for the LTE-small, NLTE, and NLTE-cool models,
and Table \ref{tabsunstats} also includes $\sigma$ values for the
comparison of the NLTE-cool $f_\lambda$ distribution to the four
data sets.  The best fit NLTE-cool model has $\Delta T_{kin}(-\infty)=-400$ K,
corresponding to a $\Delta T_{\rm kin}(\log\tau_{\rm 5000})$ offset from the NLTE model
with a slope of $\approx 70$ K in the $-\infty < \log\tau_{\rm 5000} < 0$ range.    
The NLTE-cool model is, by design, in better agreement than the NLTE model with 
data of \citet{necklabs}, although none of these models is significantly discrepant
with the observed $f_\lambda$ values.  However at longer $\lambda$ values in the
visible band the NLTE-cool model significantly under-predicts the observed
$f_\lambda$ values, as is expected for a model with a $T_{\rm kin}(\tau)$ structure
that is significantly cooler than standard solar atmospheric models.

\begin{figure}
\plotone{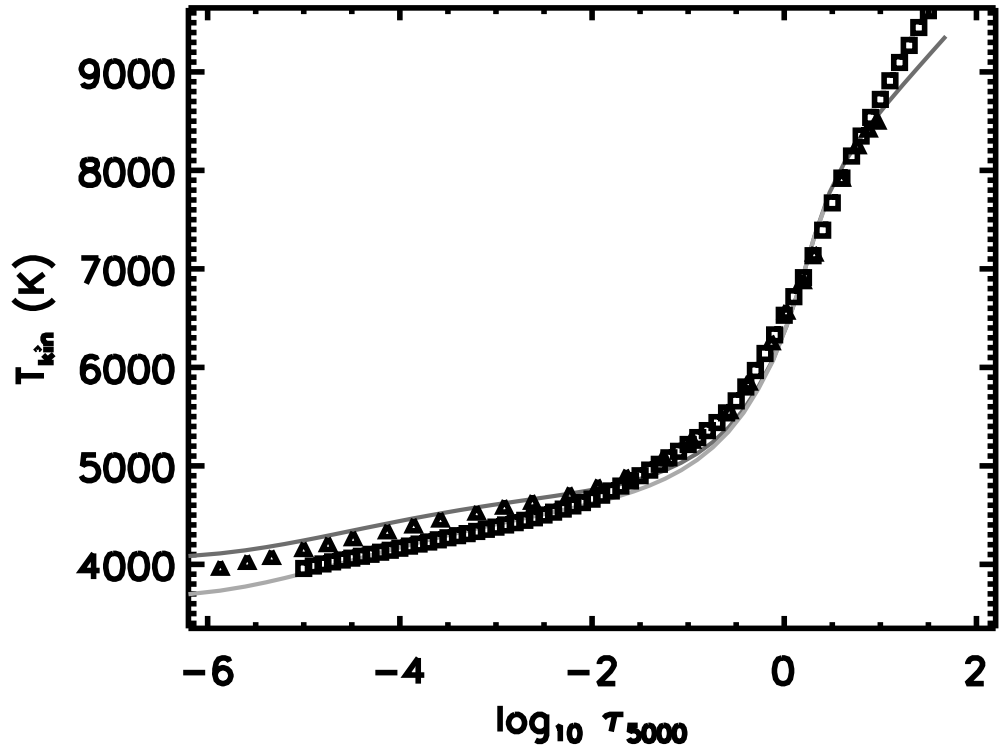}
\caption{Sun: atmospheric $T_{\rm kin}(\log\tau_{\rm 5000})$ structure.
Theoretical models: NLTE: dark line; NLTE-cool: lighter line.  
Semi-empirical models of \citet{holmul}: triangles; \citet{grev_s99}: squares.
\label{coolstruc}}
\end{figure}

\begin{figure}
\plotone{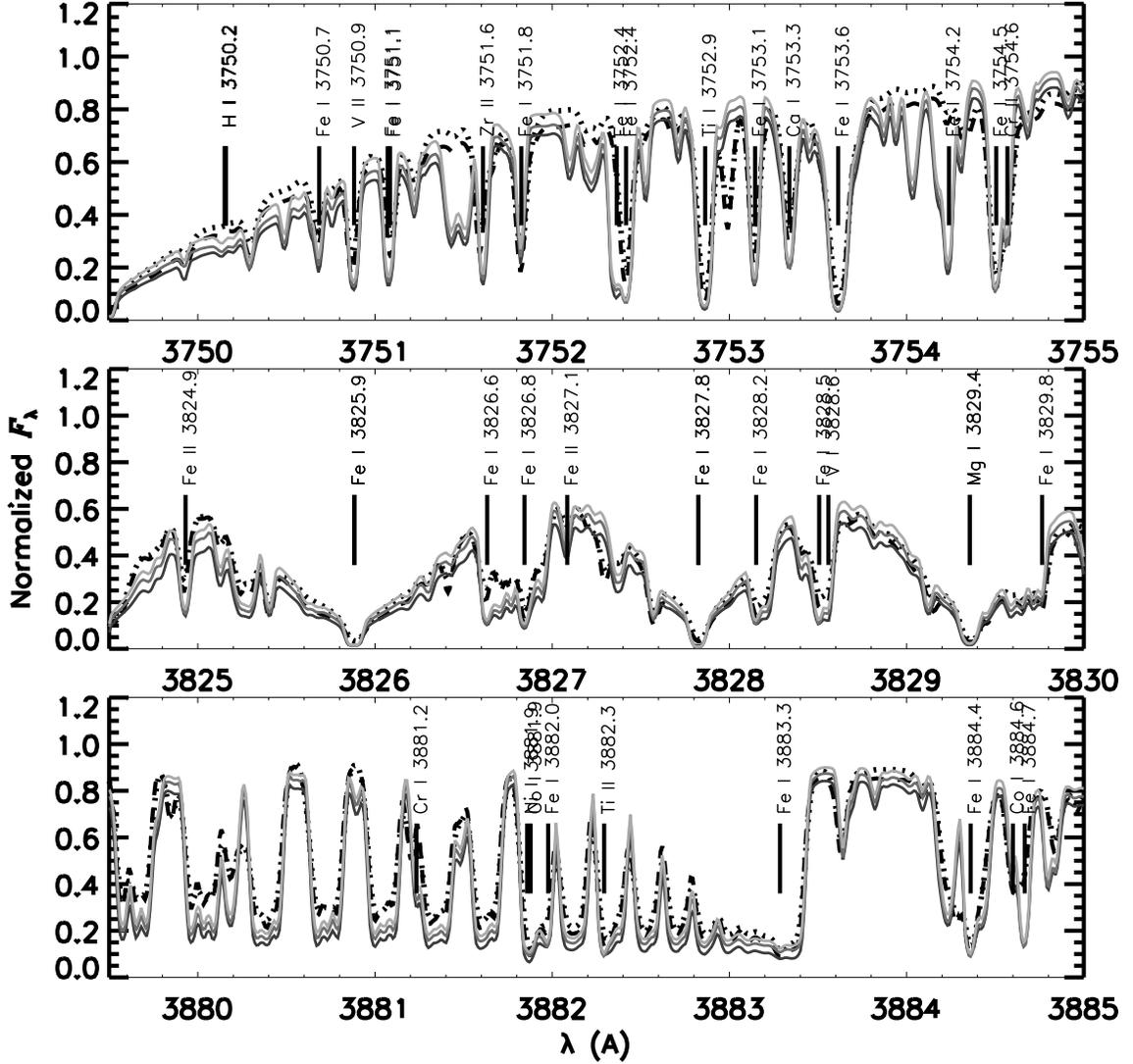}
\caption{Sun: Comparison of observed spectra of \citet{kurucz_fbt} (dashed line) and \citet{wallace} (dotted line) with computed high resolution flux
spectra ($f_\lambda(\lambda)$).  
Model LTE: dark line; model NLTE: medium line,
model NLTE-extra: light line.  
Note the relative quality of the model fits 
to the red wing of the \ion{Fe}{1} $\lambda 3749.5$ line, \ion{Mg}{1} $\lambda 3829.4$
and \ion{Fe}{1} $\lambda 3825.9$ lines, and to the CN $\lambda 3883$ band head. 
\label{sun_atlas1}}
\end{figure}

Fig. \ref{coolstruc} shows the $T_{\rm kin}(\log\tau_{\rm 5000})$ structures
of the NLTE-cool and NLTE models, and, for comparison, the $T_{\rm kin}(\log\tau_{\rm 5000})$
structures of the LTE semi-empirical solar atmospheric models of \citet{holmul} and
\citet{grev_s99}.  The model of \citet{holmul} is based on fitting the synthetic
high resolution $f_\lambda$ spectrum computed with an LTE model with an
{\it ad hoc} $T_{\rm kin}(\log\tau_{\rm 5000})$ structure to a variety of spectral 
features and to the
 center-to-limb variation of the intensity spectrum, $I_\lambda(\lambda,\cos\theta)$,
whereas the model of \citet{grev_s99} is based on an LTE analysis of the \ion{Fe}{1} 
excitation equilibrium.  Note that neither of these models show evidence of the 
well-known chromospheric $T_{\rm kin}(\log\tau)$ inversion that sets in near
$\log\tau_{\rm 5000})=-4$ because both investigators pointedly avoided spectral
features that show evidence of chromospheric influence when choosing their 
fitting diagnostics.  Interestingly, the $T_{\rm kin}(\log\tau_{\rm 5000})$ structure 
of model NLTE-cool is closer to that of \citet{holmul} than that of the NLTE
model, whereas the the NLTE model structure is closer to that of \citet{grev_s99}.

\subsection{High resolution spectroscopy}


Fig. \ref{sun_atlas1} shows the comparison between 
the observed rectified high resolution solar spectrum of \citet{kurucz_fbt} 
and synthetic spectra computed with the LTE, NLTE, and NLTE-extra models.
We have chosen three regions that demonstrate how
$f_\lambda/f_\lambda^{\rm c}$ varies when
$\kappa_\lambda/\kappa_\lambda^{\rm c}$ varies greatly over a narrow
$\Delta\lambda$ range due to a variety of different opacity sources,
such as strongly damped spectral lines from a number of chemical 
species, and molecular band heads.  The value of $f$ is small
enough that there is only s slight difference between the spectra 
of the NLTE, and NLTE-extra models, although there is marginal evidence
that the latter gives rise to slightly more flux in very strong features,
such as some strong atomic lines and the CN $\lambda 3883$ band-head.  


On balance,
the fit to high resolution spectral features does not by itself provide 
compelling evidence for the existence of missing near-UV band opacity.  
On the other hand, the addition of continuous opacity in an amount needed
to improve the fit to the observed spectrophotometry does not significantly 
worsen the fit to the high resolution spectrum.  Therefore, high resolution
spectroscopy does not rule out such missing opacity.

\section{Arcturus}

\subsection{Spectrophotometric $f_\lambda$ distribution}

\begin{figure}
\plotone{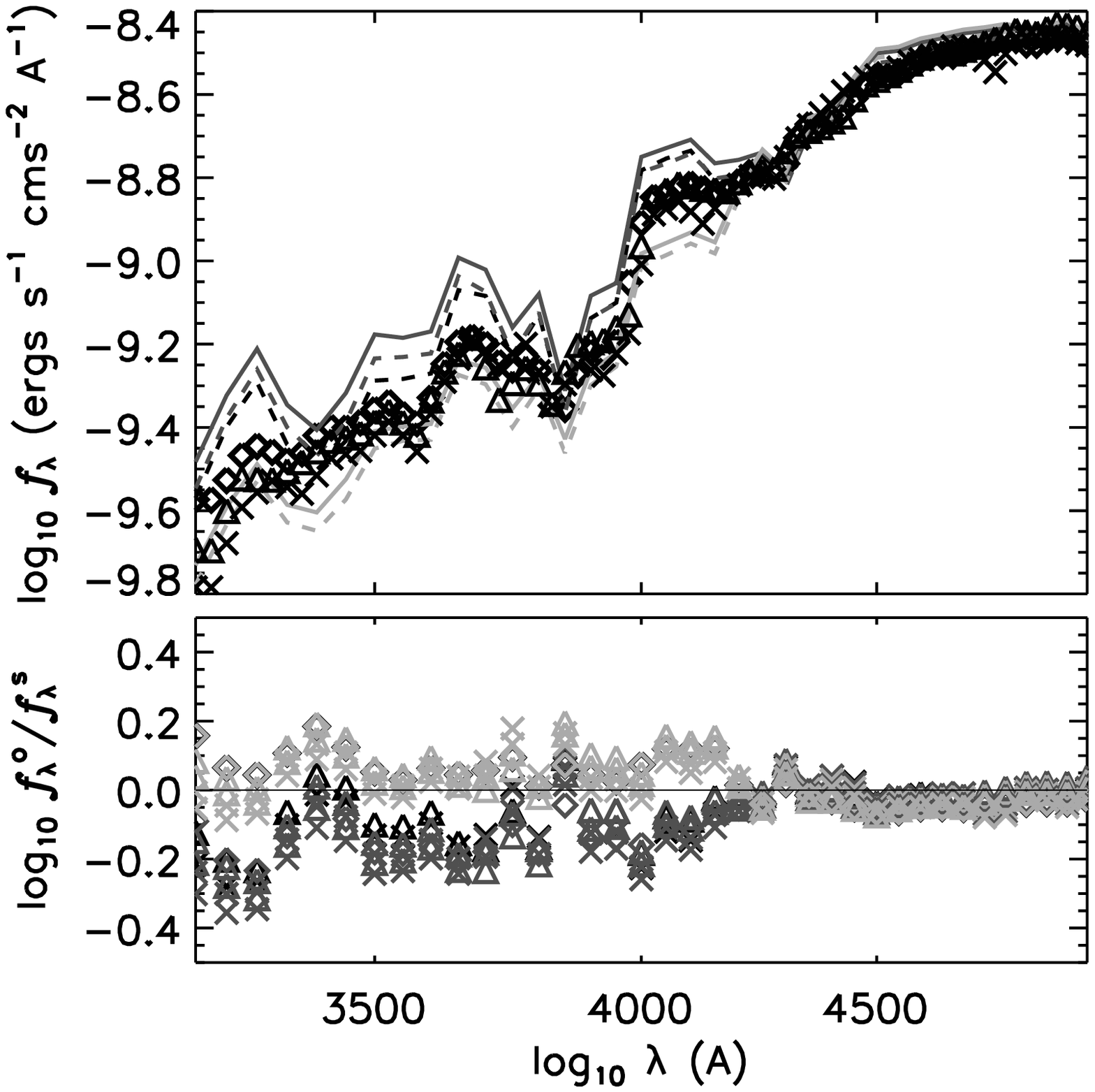}
\caption{Arcturus: Comparison of observed and computed near 
UV band logarithmic flux
distributions ($\log f_\lambda(\log\lambda)$).  
Upper panel: Direct comparison.  Lower panel: Difference between 
$f_\lambda^{\rm o}$ and $f_\lambda^{\rm s}$, 
$\log f_\lambda^{\rm o}/f_\lambda^{\rm s}$.
Observed spectra: \citet{burn}: diamonds; \citet{glushneva84}: triangles;
\citet{kharitonov78}: Xs.  
Computed spectra: LTE: dashed lines; NLTE: solid lines.  LTE-big model 
(same as LTE synthetic spectrum of Paper II \citep{shorth05}): 
black dashed line; 
LTE-small model: dark gray dashed line; LTE-small-extra model: light gray dashed line; 
NLTE model: dark gray solid line; NLTE-extra model: light gray solid line.
Note that, for simplicity, we only include 
$\log f_\lambda^{\rm o}/f_\lambda^{\rm s}$ for the NLTE and NLTE-extra models 
in the lower panel.
\label{alpboo_flxcmpuv}}
\end{figure}

Fig. \ref{alpboo_flxcmpuv} shows the comparison of the observed $f_\lambda$
distributions to the synthetic spectra computed with models that explore
the variation in the treatment of the thermodynamic equilibrium and the
value of $\kappa_\lambda^{\rm c}$.  Table \ref{tabarctstats} presents
the $\sigma$ values for each comparison computed as described in Section 
\ref{sunabsolut}.  The synthetic surface flux spectra,
$F_\lambda$, have been geometrically diluted to the flux at Earth,
$f_\lambda$, with an angular diameter
for Arcturus, $\phi$, of 21.0 mas (\citep{griffinl99}), which, as demonstrated
in Paper I, provides a good match to the visible and near IR observed
$f_\lambda$ distribution.  As reported in Paper I, both
the LTE-big and NLTE models predict too much flux in the near UV band
as compared to all four observed $f_\lambda$ distributions, 
although the NLTE model is more discrepant with the data than the LTE-big model.
As with the Sun, to remove the choice of atomic line list as a variable among 
the models, we have recomputed the LTE model of Paper I with the ``small''
line list that the NLTE model also necessarily uses, producing our LTE-small
model for Arcturus.  As expected, this model is less discrepant with the 
NLTE model, but, as a result, is even more discrepant with the observed values.
As with the Sun, we note that in Paper I we have over-estimated the
amount of NLTE UV band $f_\lambda$ brightening by using inconsistent
line lists in the LTE and NLTE modeling.

\paragraph{}

The need for some adjustment to the model that reduces the UV band flux 
is even more obvious for Arcturus than for the Sun.  We have computed a
new NLTE model for Arcturus, NLTE-extra, in which the value of $f$ is
unity, {\it ie.} the value of $\kappa_\lambda^{\rm c, T}$ is double
that of a standard model.  This model provides a close fit to the overall
observed $f_\lambda$ level, although the detailed distribution deviates 
locally from the observed one, most notably around 3400 \AA~ and 4100 \AA.
NLTE-extra represents a large amount of missing thermal absorption
opacity in the near UV band of early K giants.   We note that the amount
of extra continuous thermal extinction that we require to force
NLTE models to agree with observed near UV band spectrophotometry is
consistent with the results of \citet{short94}, who found on the 
basis of LTE modeling with the ATLAS9 code that $\kappa_\lambda^{\rm c, T}$
had to approximately double. 

\paragraph{}

The LTE-small-extra spectrum consistently lies just below the NLTE-extra
spectrum throughout the 3000 to 4200 \AA~ region, and, as with the Sun, shows a 
relation to the NLTE-extra spectrum that is similar to that of the LTE-small spectrum to 
the NLTE spectrum.  Again, without performing a detailed analysis of the
effect of $f\times\kappa_\lambda^{\rm c, T}$ on the NLTE model, we conclude to a first 
approximation that the extra extinction does not change the overall character of the 
NLTE effects on $f_\lambda$.  We note that, in contrast to the solar case,
the LTE-small-extra spectrum lies substantially below that of the LTE-big
model.  For Arcturus, the value of $f$ is so large that it more than compensates
for the extra extinction provided by the extra lines in the ``big'' line list.  
As a result, it is much more clearly the case with Arcturus than with the Sun that 
the inability
of standard models to reproduce the observed near UV flux cannot be addressed
by choice of line list.   

\paragraph{}

\citet{pdk93} concluded that Arcturus has a non-solar abundance distribution
with some $\alpha$-process elements being enhanced by about a factor of 
two with respect to the overall scaled solar value.  We computed NLTE
atmospheric models and spectra with the abundances of \citet{pdk93} and
found that while there were detectable differences among the 
low resolution $f_\lambda$ distributions computed with the 
scaled solar and non-solar models, they were much smaller than the 
the discrepancy between the observed and computed spectra without
$f\times\kappa_\lambda^{\rm c, T}$, and also much smaller than the difference
between the flux spectra computed with LTE and equivalent NLTE models.
We conclude that the near UV band $f_\lambda$ discrepancy in Arcturus
is not caused by inaccurate model abundances.

\paragraph{}

\begin{figure}
\plotone{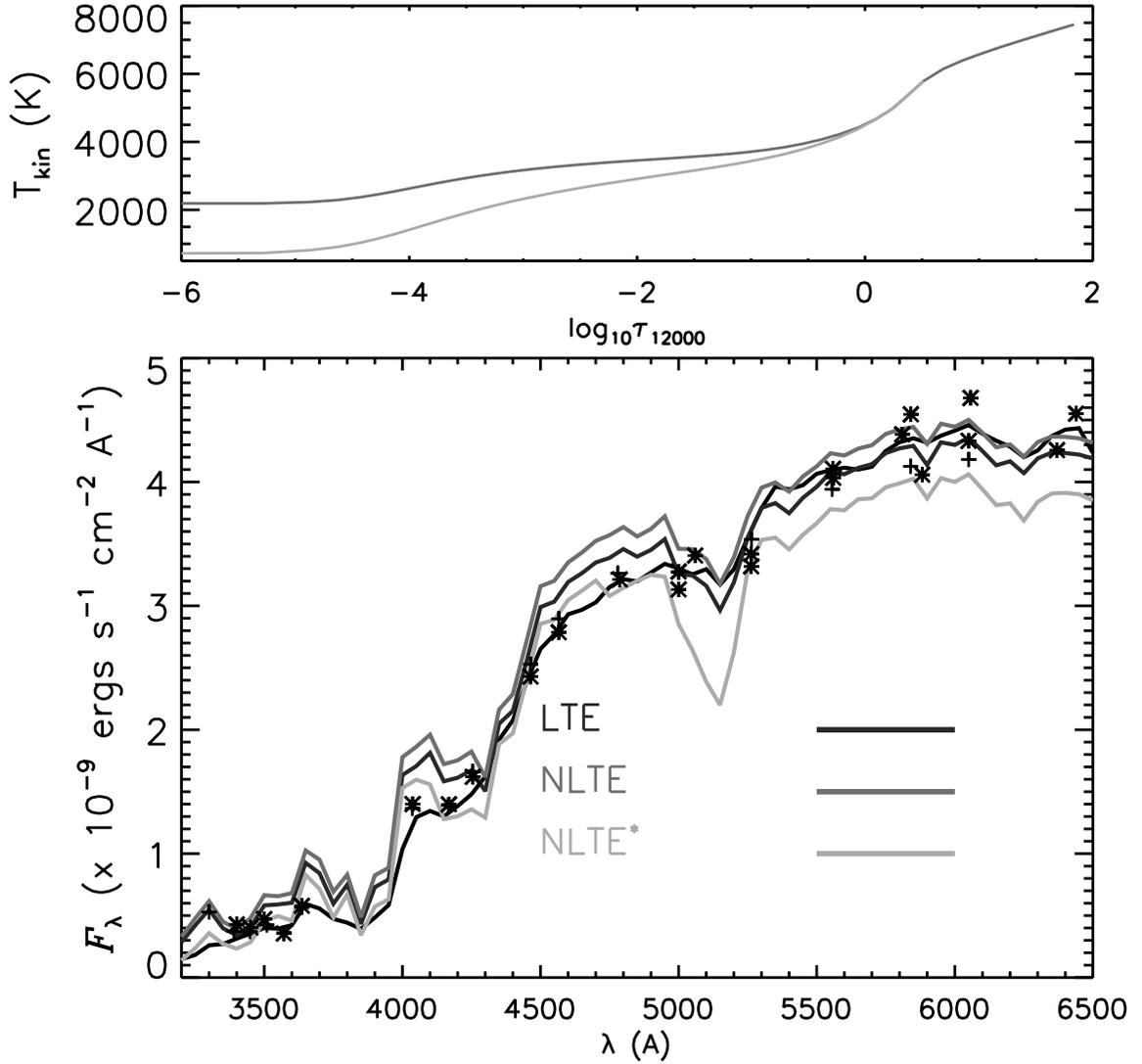}
\caption{Arcturus: Upper panel: Atmospheric $T_{\rm kin}(\log\tau_{\rm 5000})$ structure.
Theoretical models: NLTE: dark line; NLTE-cool: lighter line.
Lower panel: Same as Fig. \ref{alpboo_flxcmpuv}, except that the models explore
the variation in $T_{\rm kin}(\tau_{\rm 5000})$ structure.  LTE-small:
darkest color; NLTE model: medium gray color; NLTE-cool model: lightest color.
\label{alpboo_cool}}
\end{figure}

As with the Sun, we attempted to construct a model, NLTE-cool, for Arcturus,
that improves the fit to the observed near UV band $f_\lambda$ values
by {\it ad hoc} adjustment of the $T_{\rm kin}(\log\tau)$ structure.
We were able to investigate models with $\Delta T_{kin}(-\infty)$ as large as -1500 K,
corresponding to a $\Delta T_{\rm kin}(\log\tau_{\rm 5000})$ offset from the NLTE model
with a slope of $\approx 250$ K in the $-\infty < \log\tau_{\rm 5000} < 0$ range.    
The $\Delta T_{kin}(-\infty)=-1500$ K model, shown in Fig. \ref{alpboo_cool}, has 
an upper boundary 
temperature at $\log\tau_{\rm 5000}=-6$ of only $\approx 500$ K, and models with
larger $\Delta T_{kin}(-\infty)$ values are numerically unstable because 
the hydrostatic equilibrium solution has negative pressures at some depths.
Therefore, the model of $\Delta T_{kin}(-\infty)=-1500$ K represents the most
extreme $T_{\rm kin}(\log\tau)$ structure that we were able to produce, and even it
does not reduce the predicted $f_\lambda$ values in the near UV to the observed ones,
as can be seen in Fig. \ref{alpboo_cool}.  We conclude that of the two
modifications to the model proposed here, an additional source of
$\kappa_\lambda^{\rm c}$ can more effectively address the discrepancy with
observations.
Moreover, model NLTE-cool predicts
significantly too little flux in the well observed visible band.  We note that
model NLTE-cool represents the extreme limit of so-called 1.5D modeling of horizontal
thermal inhomogeneity in the atmosphere of Arcturus with components that all have
$T_{\rm eff}\approx 4250$ K.  A semi-empirical 
component that is significantly cooler than an RE 
model is still unable to reproduce the observed near UV flux, even when this cool component
has a weight of unity in the ``1.5D model''.  Therefore, we conclude that 1.5D models
with a mixture of hotter and cooler components that all have $T_{\rm eff}\approx 4250$ K cannot 
resolve the near UV discrepancy for Arcturus. 
 
\paragraph{}

We note that previous authors have also investigated non-radiative cooling in the 
outer atmosphere of Arcturus to address other observational discrepancies.
\citet{ryde} observed H$_{\rm 2}$O vapour lines in the mid-IR spectrum of Arcturus
and found that the $T_{\rm kin}(\log\tau)$ structure computed with a MARCS model in 
RE had to be depressed {\it ad hoc} by $\approx 300$ K
in the $\log\tau_{\rm 5000} < -3.8$ range.  However, the model of \citet{ryde} has a 
boundary temperature of $\approx 1800$ K, 1300 K warmer than our NLTE-cool model, and
comes into agreement with the RE model at $\log\tau_{\rm 5000} = -3.8$, at which
depth our NLTE-cool model is still $\approx 1000$ K cooler than our RE model.   
\citet{wiedemann} constructed a semi-empirical $T_{\rm kin}(\log\tau)$ structure 
to account for the observed central flux of the CO $\Delta\nu=1$ near-IR lines.  
They compared this model to a semi-empirical structure with a 
chromospheric $T_{\rm kin}(\log\tau)$ inversion fitted to chromospheric spectral
diagnostics rather than to an RE model.  The surprising depth of the CO $\Delta\nu=1$
lines seems to demand a significant alteration to horizontally homogeneous 
semi-empirical models based on visible and near UV spectral features, yet, even the
CO based semi-empirical model of \citet{wiedemann} has a boundary temperature of
$\approx 2400$ K at a depth of logarithmic column mass density of $-2.0$ (cgs units).  
This column mass density corresponds to a $\log\tau_{\rm 5000}$ value of 
$\approx -5.9$ in our model where the $T_{\rm kin}$ value is less than 800 K, which, 
again, is over 1200 K cooler than the \citet{wiedemann} model.  That 
our NLTE-cool model is so extreme compared to other semi-empirical models with cooled
outer atmospheres based on other spectral diagnostics is further evidence that this 
model is unrealistic.

\subsection{High resolution spectroscopy}

\begin{figure}
\plotone{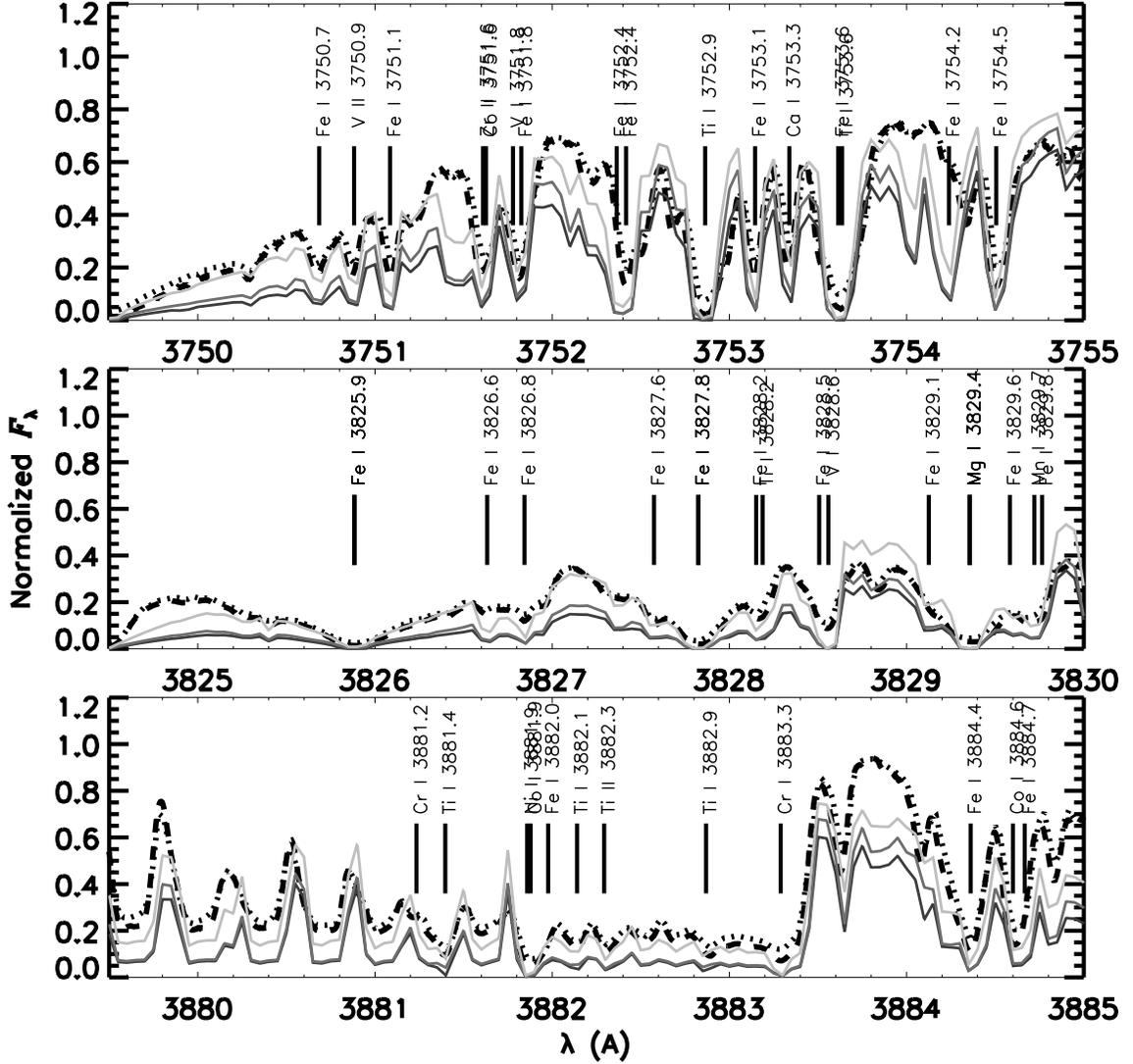}
\caption{Arcturus: Comparison of observed and computed high resolution flux
spectra ($f_\lambda(\lambda)$).  Observed spectra: \citet{hinkle_wvh00}: dashed
line.
Model LTE: dark line; model NLTE: medium line, model NLTE-extra: light line.
dashed line; \citet{griffin68}: dotted line.  
Note the relative quality of the model fits 
to the red wing of the \ion{Fe}{1} $\lambda 3749.5$ line, \ion{Mg}{1} $\lambda 3829.4$
and \ion{Fe}{1} $\lambda 3825.9$ lines, and to the CN $\lambda 3883$ band head. 
\label{alpboo_atlas1}}
\end{figure}

Fig. \ref{alpboo_atlas1} shows the comparison 
between the observed high resolution spectra of both \citet{hinkle_wvh00} and
\citet{griffin68},
 and synthetic spectra computed with our LTE, NLTE, and NLTE-extra models
for the same spectral regions that were shown for the fit to the Sun.
As with the Sun, the strongly damped \ion{Fe}{1} lines have slightly brighter wings
as a result of NLTE \ion{Fe}{1} over-ionization.  However, again, the difference
is almost negligible for many lines.  By contrast with the Sun, the NLTE-extra model
produces strong broad spectral features that are significantly brighter, and in
much better agreement with the observed spectrum, than either the LTE or NLTE models.
This agreement can be seen for strongly damped lines of both \ion{Fe}{1}, \ion{Mg}{1},
and the CN $\lambda3883$ band head.  It is significant that we re-affirm with NLTE
modeling the conclusion that \citet{short96} drew with LTE modeling: namely, that
the same enhancement of $\kappa_\lambda^{\rm c}$ that brings the near UV absolute 
$f_\lambda$ levels into agreement with the observed values, and also simultaneously
brings the strong, broad features in the near UV high resolution spectrum into 
better agreement with data.

\section{Procyon}

\subsection{Spectrophotometric $f_\lambda$ distribution}

\begin{figure}
\epsscale{.80}
\plotone{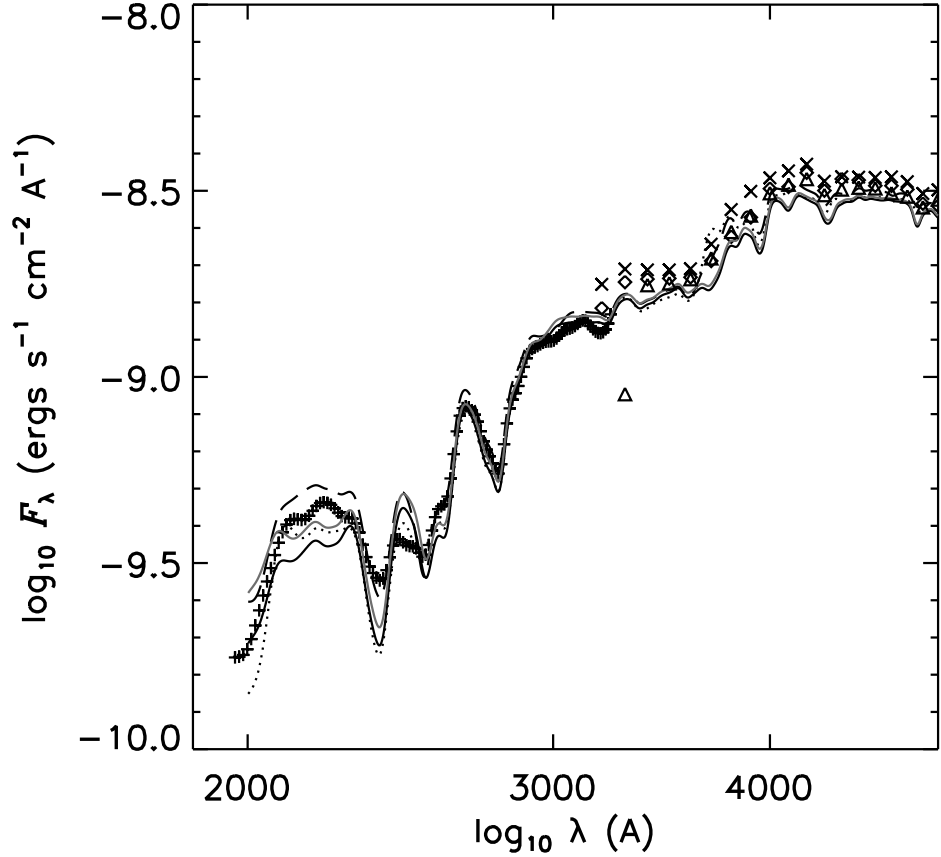}
\caption{Procyon: Comparison of observed and computed near 
UV band logarithmic flux
distributions ($\log f_\lambda(\log\lambda)$).  
Observed spectra; of \citet{burn}: diamonds; \citet{glushneva84}: triangles;
\citet{kharitonov78}, IUE: crosses.  
Computed spectra: LTE-big model: darkest line; 
LTE-small model: dark gray; NLTE model: light gray; NLTE-extra
model: lightest color.
\label{proc_flxcmpuv}}
\end{figure}

In Fig. \ref{proc_flxcmpuv} we present the comparison between observed
and modeled $f_\lambda$ distributions for Procyon.  The synthetic
$F_\lambda$ spectra have been de-projected to $f_\lambda$ values at Earth
using a value of $\phi$ of 5.404 mas \citet{jaufdenberg}.  The NLTE models
agree more closely with the LTE models for Procyon than they do for
the Sun or Arcturus, and, by contrast with the results for the Sun and 
Arcturus, both the LTE and NLTE models match the observed $f_\lambda$
distribution well throughout the near to mid UV band, including the 
range observed by IUE.  If anything, there is marginal evidence that 
the models predict too little flux in the near UV band.  We find no evidence
for the need for extra extinction, nor any other adjustment to the 
modeling, in the case of the hottest program star.  We conclude that 
by $T_{\rm eff}$ values of about that of the Sun, or slightly hotter, 
the near UV extinction is well reproduced by models, and do not 
present further discussion of Procyon here.

%
%
%

\section{Extinction sources}

\begin{figure}
\plotone{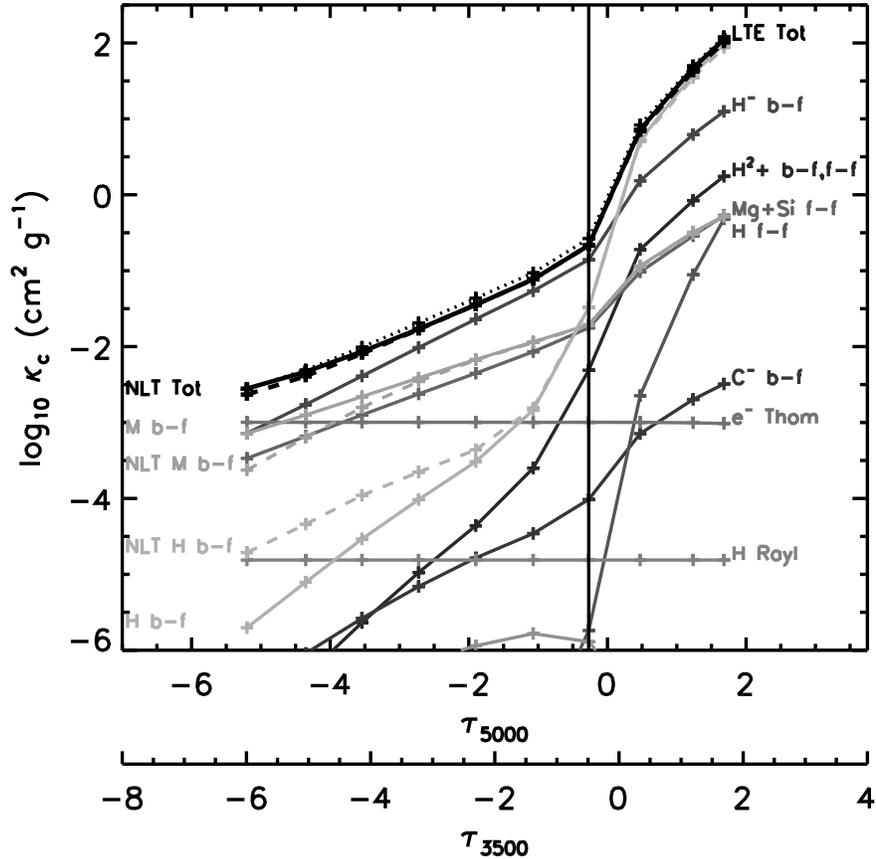}
\caption{Sun: Depth dependence of the contribution to the total monochromatic
 continuous extinction, $\kappa_\lambda^{\rm c}$, at $\lambda = 3500$ \AA~ from 
all significant sources.  
To limit the size of output files, only every tenth
depth point in our 50 depth point model is plotted.  Extinctions are computed
in LTE (solid lines) and NLTE (dashed lines).  (Note that only two of the major 
contributors,
\ion{H}{1} {\it b-f} and ``metal'' {\it b-f}, are treatable in both LTE and 
NLTE with 
PHOENIX (see text)).  The dotted line shows the total extinction with the
addition of the {\it ad hoc} extra continuum extinction, 
$\kappa_\lambda^{\rm c}+f\times \kappa_\lambda^{\rm c, T}$ (see text).  
The vertical line indicates the $\tau_{\rm 5000}$ value 
at which the
$\lambda$ dependence of $\kappa_\lambda^{\rm c}$ is plotted in Figs.
\ref{opcsunwave} and \ref{opcsunwave2}. 
 \label{opcsundepth}}
\end{figure}

\begin{figure}
\plotone{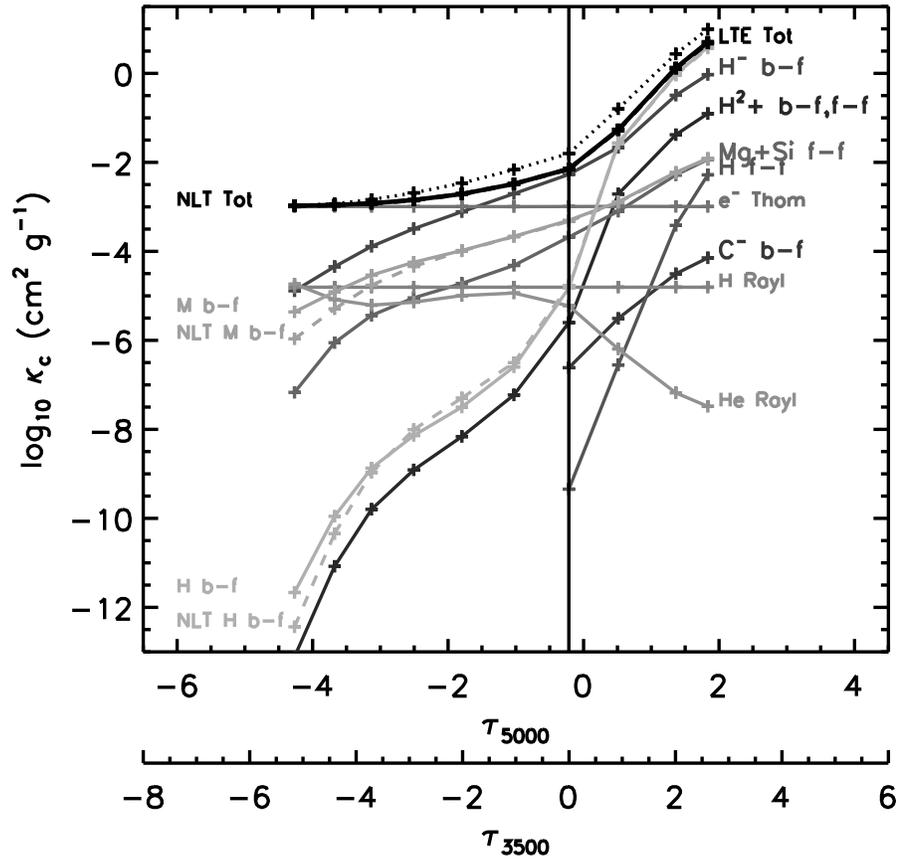}
\caption{Arcturus $\kappa_\lambda^{\rm c}(\tau)$: Same as Fig. \ref{opcsundepth}, except for Arcturus. 
The vertical line indicates the $\tau_{\rm 5000}$ value 
at which the
$\lambda$ dependence of $\kappa_{\rm c}, \lambda$ is plotted in Figs.
\ref{opcalpboowave} and \ref{opcalpboowave2}. 
 \label{opcalpboodepth}}
\end{figure}

\begin{figure}
\plotone{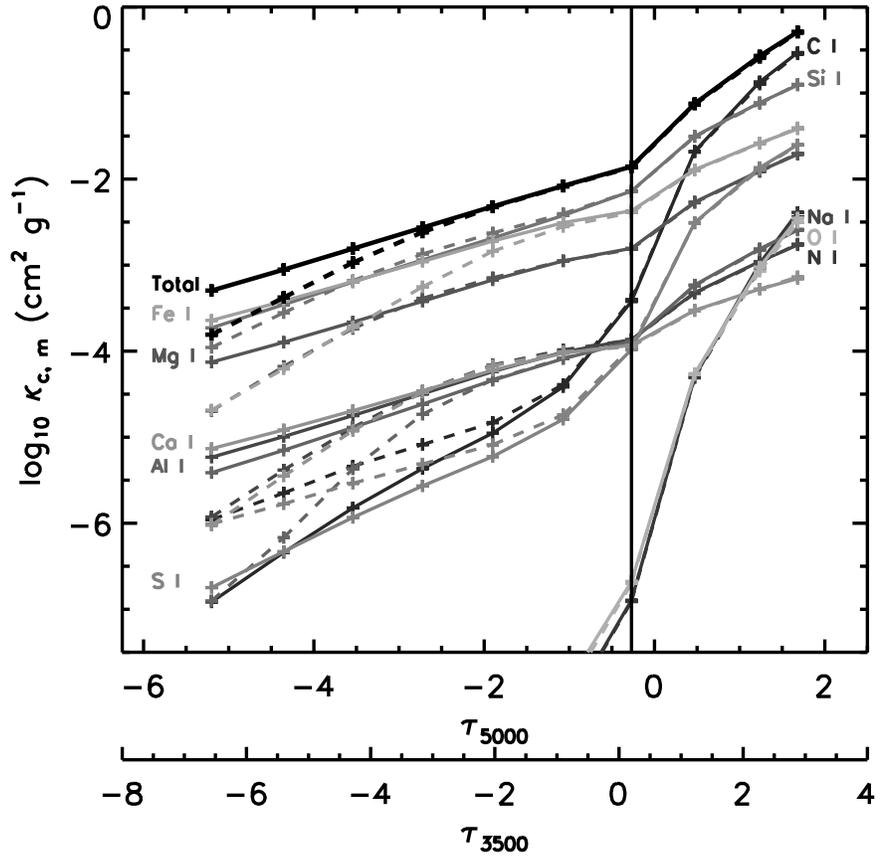}
\caption{Sun $\kappa_\lambda^{\rm c, m}(\tau)$: Same as Fig. \ref{opcsundepth}, except that the contribution to the 
total monochromatic
 continuous ``metal'' {\it b-f} extinction, $\kappa_\lambda^{\rm c, m}$, 
from all significant metals is shown.
 \label{opcsundepth2}}
\end{figure}

\begin{figure}
\plotone{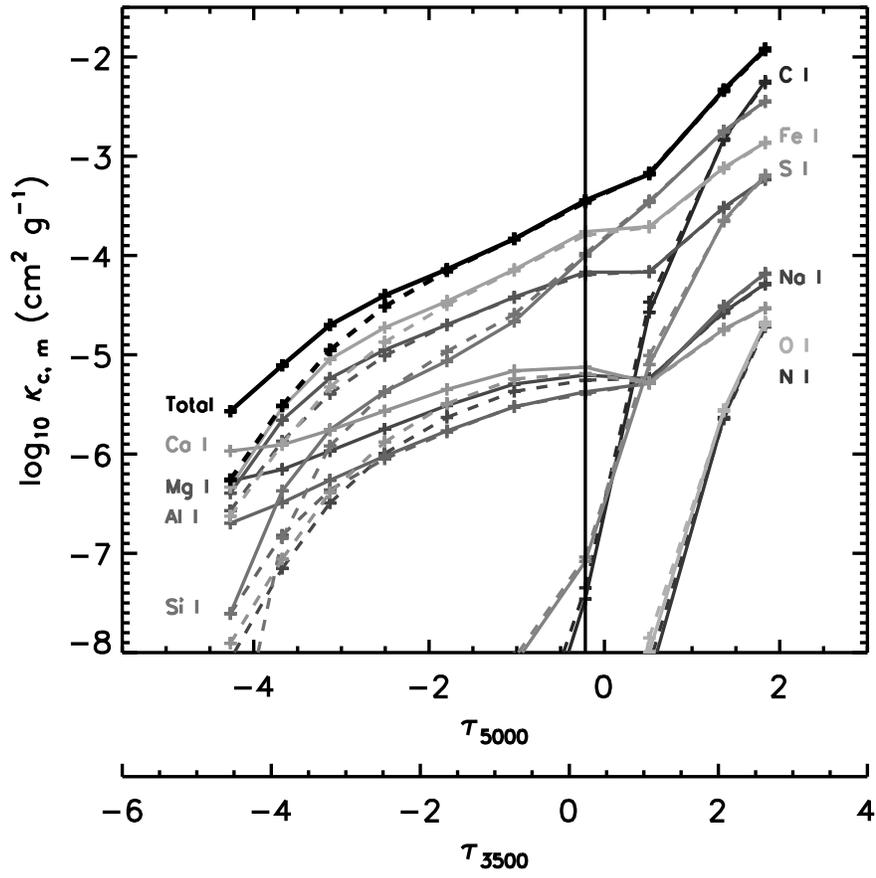}
\caption{Arcturus $\kappa_\lambda^{\rm c, m}(\tau)$: Same as Fig. \ref{opcsundepth2}, except for Arcturus.
 \label{opcalpboodepth2}}
\end{figure}

\begin{figure}
\plotone{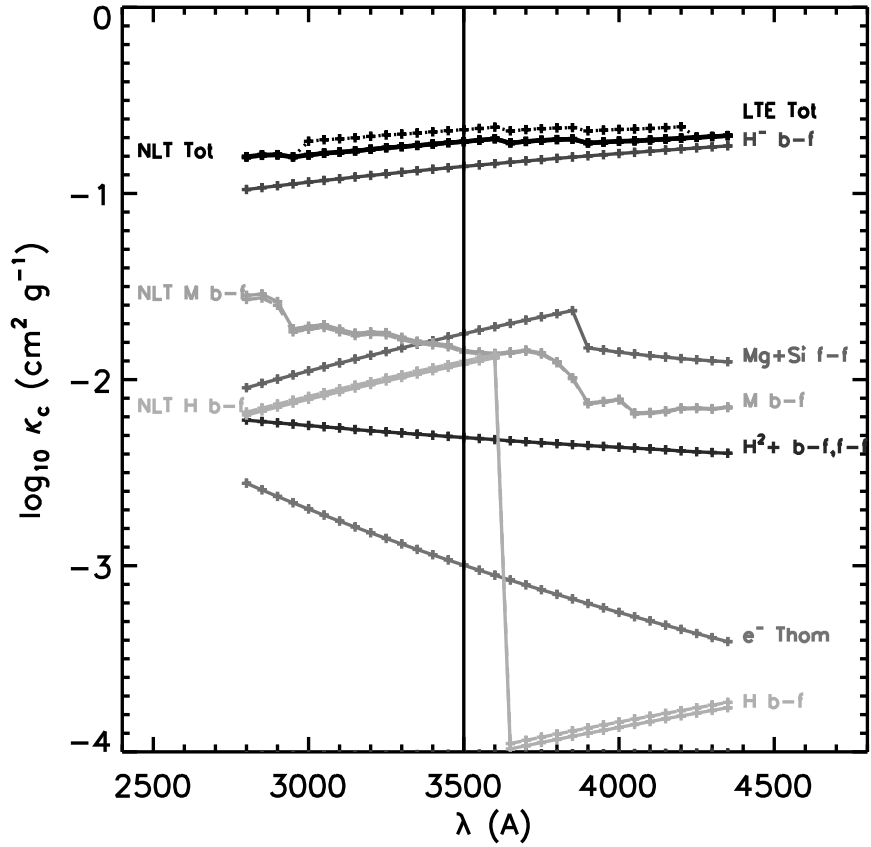}
\caption{Sun $\kappa_\lambda^{\rm c}(\lambda)$: Same as Fig. \ref{opcsundepth}, except that the $\lambda$ dependence 
in the near-UV band of $\kappa_\lambda^{\rm c}$, at $\tau_{\rm 5000}=0.3$
 from all significant sources is shown.  The vertical line indicates the 
$\lambda$ value at which the depth dependence of $\kappa_\lambda^{\rm c}$ 
is shown in Figs. \ref{opcsundepth} and \ref{opcsundepth2}. 
 \label{opcsunwave}}
\end{figure}

\begin{figure}
\plotone{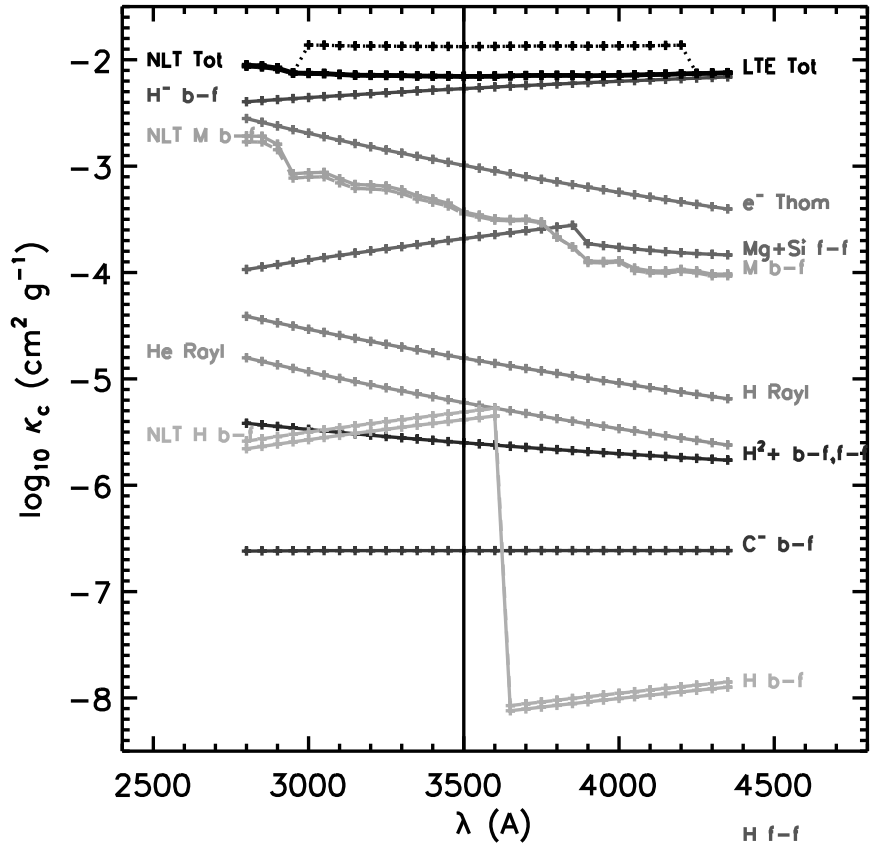}
\caption{Arcturus $\kappa_\lambda^{\rm c}(\lambda)$: Same as Fig. \ref{opcsunwave}, except for Arcturus.
The vertical line indicates the 
$\lambda$ value at which the depth dependence of $\kappa_{\rm c}, \lambda$ 
is shown in Figs. \ref{opcalpboodepth} and \ref{opcalpboodepth2}. 
 \label{opcalpboowave}}
\end{figure}

\begin{figure}
\plotone{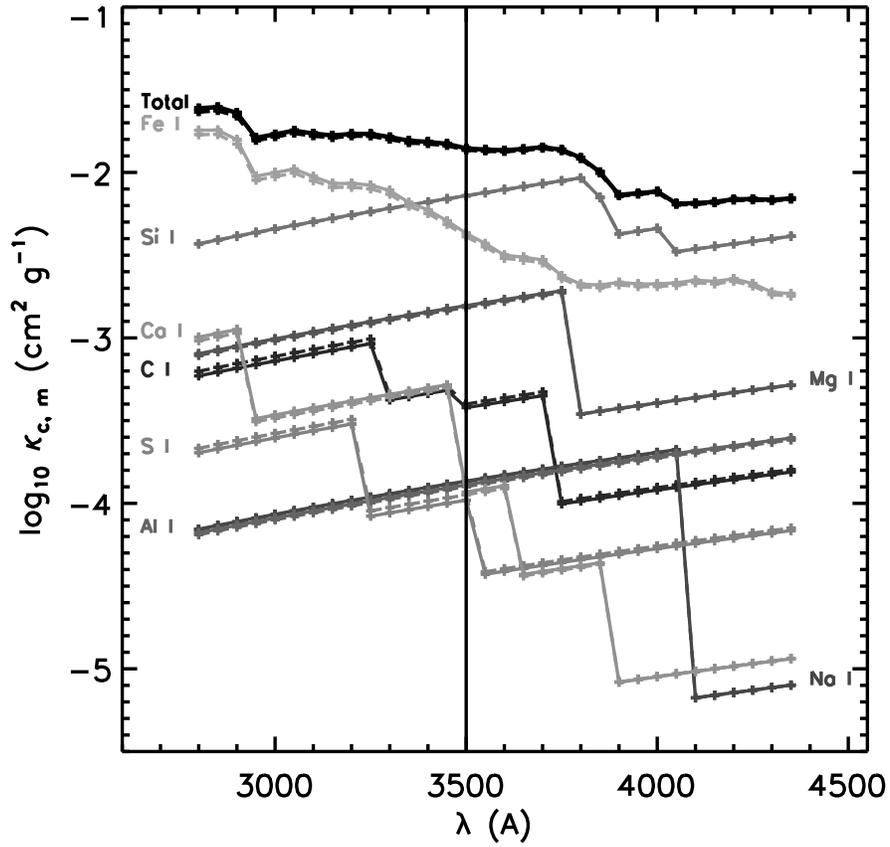}
\caption{Sun $\kappa_\lambda^{\rm c, m}(\lambda)$: Same as Fig. \ref{opcsundepth2}, except that the $\lambda$ dependence 
in the near-UV band of the contribution to $\kappa_\lambda^{\rm c, m}$, 
at $\tau_{\rm 5000}=0.3$ from all significant metals is shown.
 \label{opcsunwave2}}
\end{figure}

\begin{figure}
\plotone{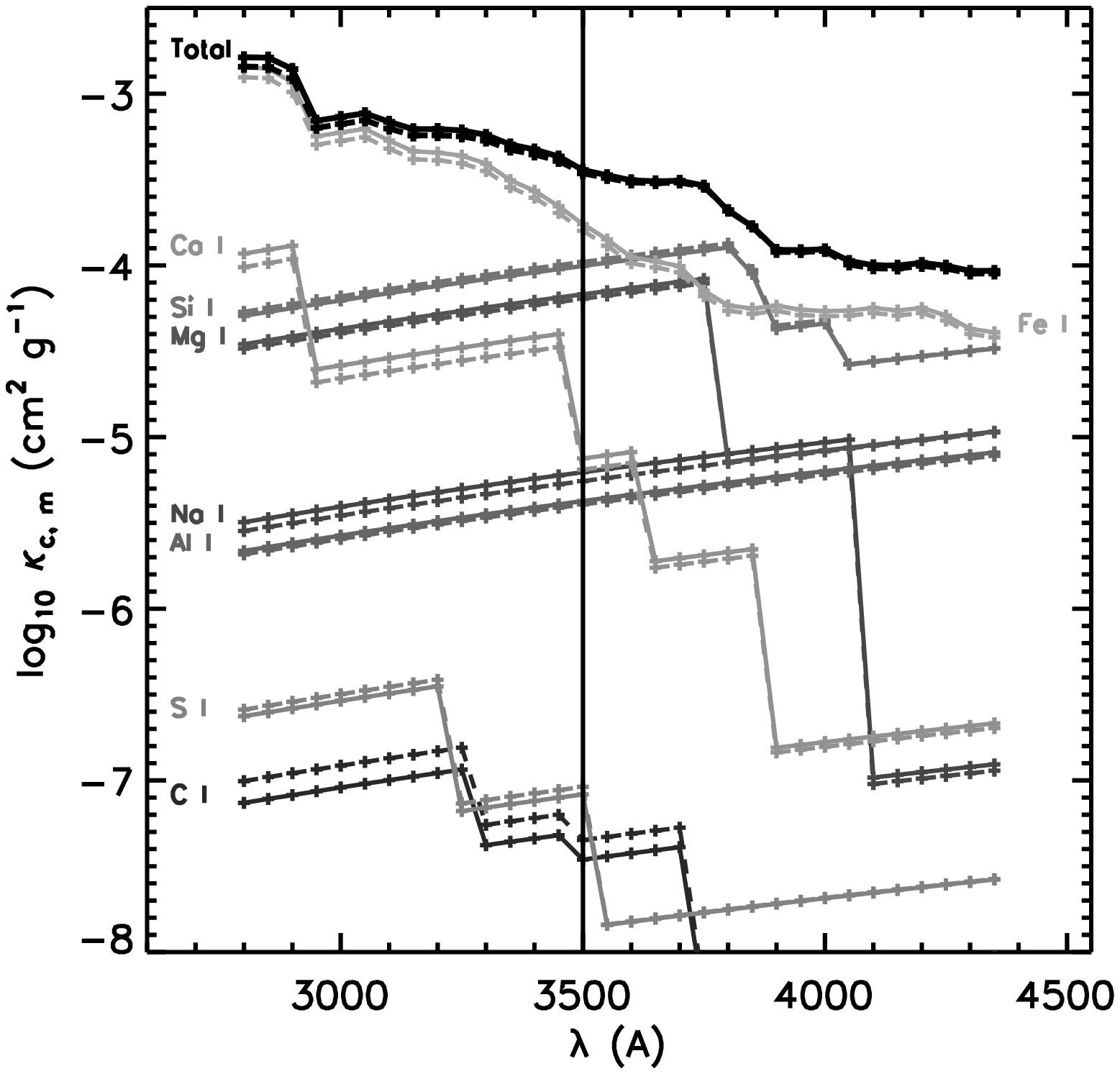}
\caption{Arcturus $\kappa_\lambda^{\rm c, m}(\lambda)$: Same as Fig. \ref{opcsunwave2}, except for Arcturus.
 \label{opcalpboowave2}}
\end{figure}

Figs. \ref{opcsundepth} and \ref{opcalpboodepth} show the dependence on the 
standard depth variable, $\log\tau_{\rm 5000}$, of the most important thermal 
absorption and scattering contributions to
$\kappa_\lambda^{\rm c}$ for the Sun and Arcturus, respectively, at a $\lambda$ 
value of 3500 \AA, near the center
of the band for which we are proposing an additional continuous extinction 
source.  
We also include the $\log\tau_{\rm 3500}$ scale among the abscissae 
so that an assessment of the value of $\kappa_\lambda^{\rm c}$ near
monochromatic optical depth unity can be made at the relevant wavelength. 
However, the two scales do not differ greatly for any star.  For
both $\tau_\lambda$ scales, we present continuum {\it extinction} optical depth,
due to both thermal absorption and scattering extinction. 
  To control the size of extinction co-efficient output files,
we have only tabulated and plotted the $\kappa_\lambda$ values for ten 
sample depth points spaced approximately equally at intervals of $\Delta\log\tau=0.8$.  
To simplify these figures, we have presented the total bound-free ({\it b-f})
thermal extinction due to all included metals, $\kappa_\lambda^{\rm c, m}$
(labeled ``M b-f'').  In Figs. \ref{opcsundepth2}, \ref{opcalpboodepth2}
 we show the $\log\tau_{\rm 5000}$ 
dependence of the contribution of the most important individual metallic species 
to $\kappa_\lambda^{\rm c, m}$ at the same $\lambda$ value for the two stars.  
Figs. \ref{opcsunwave} and \ref{opcalpboowave} and Figs. 
\ref{opcsunwave2} and \ref{opcalpboowave2} show the variation of 
the same quantities with $\lambda$
for a $\log\tau_{\rm 5000}$ value of $0.3$, close to where the near UV continuum flux 
arises, with a sampling, $\Delta\lambda$, of
25 \AA.  In all cases, we show the results of both LTE and NLTE calculations
for those $\kappa_\lambda$ sources that can be computed in NLTE by PHOENIX.  
We note
that this is the first presentation of continuous extinction by source 
for a solar or late-type star, computed with thousands of transitions of all 
astrophysically important light 
metals and Fe-group elements treated in direct multi-level NLTE.

As expected, for the Sun and Arcturus, both the LTE and NLTE PHOENIX calculations 
confirm the long
known result that the near UV band $\kappa_\lambda^{\rm c}$ values are 
overwhelmingly dominated by H$^-$ $b-f$ extinction near continuum optical
depth unity.  
For both the Sun and Arcturus, the \ion{H}{1} $b-f$ extinction just 
begins to rival that of H$^-$ $b-f$
at the deepest layer of the model ($\log\tau_{\rm 5000}>1$). 
For the Sun the total metal $b-f$ extinction, when computed in LTE, and 
Thomson scattering both rival the \ion{H}{1} $b-f$ 
$\kappa_{\rm 3500}$ value at the highest layers of the 
model ($\log\tau_{\rm 5000}<-5$), while for Arcturus, Thomson scattering
becomes dominant in the outer atmosphere for $\log\tau_{\rm 5000}<-2$.
However, these layers are not expected to 
contribute significantly to the emergent surface flux distribution 
($F_\lambda(\tau=0)$) in the near UV pseudo-continuum.  The H$^-$ $b-f$ 
process is a well understood and accurately computable $\kappa_\lambda$ 
source, and there seems to be little scope for belief that its value
is erroneous by a large enough factor to explain the over-prediction
of near UV flux by models.

For those $\kappa_\lambda$ sources that are treatable in NLTE with PHOENIX,
we note that, for the Sun, NLTE effects depress the value of the total metal $b-f$ 
extinction at 3500 \AA~ for $\log\tau_{\rm 5000} < -2$ by as much as 0.5 dex by
$\tau_{\rm 5000}\approx 0$, and increase the value of the \ion{H}{1}
$b-f$ extinction at 3500 \AA~ for $\log\tau_{\rm 5000} < -2$ by as much as almost one dex 
by $\tau_{\rm 5000}\approx 0$.  
  For Arcturus, the NLTE effects on both the total
metal $b-f$ and H $b-f$ $\kappa_{\rm 3500}$ values are similar to that for the Sun,
except that the NLTE depression of the H $b-f$ extinction begins further out
in the atmosphere ($\log\tau_{\rm 5000} < -3$).  All NLTE $\kappa_\lambda$ values
approach the corresponding LTE value near the bottom of the atmosphere, as expected.

Figs. \ref{opcsundepth2} and \ref{opcalpboodepth2}, 
and \ref{opcsunwave2} and \ref{opcalpboowave2}, show that the NLTE depression of
the total metal $b-f$ extinction in the outer atmosphere is driven by a corresponding
depression in the $b-f$ extinction of many, but not all, of the most important individual 
metallic $\kappa_\lambda$ contributors. 
For the Sun, these are \ion{Si}{1}, \ion{C}{1}, and \ion{Fe}{1} for $\tau_{\rm 5000} > 1$.  
For $\tau_{\rm 5000} < 1$, they are \ion{Fe}{1} and \ion{Si}{1} for an LTE calculation, 
whereas, for a NLTE calculation, \ion{Si}{1} is still important, but \ion{Mg}{1} rivals
\ion{Fe}{1} due to the large drop in \ion{Fe}{1} extinction in the outer atmosphere
as a result of the well-known NLTE \ion{Fe}{1} over-ionization. 
For Arcturus, the important metallic contributors are similar to those for the Sun,
except that \ion{C}{1} is not important at large depth, and both an LTE and a NLTE 
calculation show that \ion{Fe}{1} and \ion{Mg}{1} are important in the outer atmosphere,
whereas \ion{Si}{1} is not.  
Paper I and Paper II contain a detailed discussion
of NLTE effects on the extinction of metallic elements for Arcturus and the Sun,
respectively.  However, for both stars, the effect on the total 
continuum extinction is almost negligible.  Therefore, it does not seem likely 
that the over-prediction of near UV flux is due to the inaccurate or
incomplete treatment of NLTE effects in {\it continuum} extinction
sources (the effect of NLTE on {\it line} ($b-b$) extinction, especially
that of \ion{Fe}{1}, is very significant, as described at great length in 
earlier literature (see Paper II and references therein), but we are
only investigating the role of continuum extinction here). 
However, as noted above, for none of our models does metallic $b-f$ extinction
dominate the near UV total $\kappa_\lambda^{\rm c}$ value, so, although these
NLTE results for continuum metal extinction may be interesting in their own
right, it seems unlikely that inadequacies in their NLTE modeling will be able to 
account for the failure of NLTE models to correctly predict the observed near UV
$f_\lambda$ level in the Sun and Arcturus.

Figs. \ref{opcsundepth} and \ref{opcsunwave} show the $\log\tau_{\rm 5000}$ dependence
at $\lambda=3500$ \AA~, and the $\lambda$ dependence at $\log\tau_{\rm 5000}=0.3$, 
respectively, of the standard continuous extinction, $\kappa_\lambda^{\rm c}$, and the 
enhanced value, $\kappa_\lambda^{\rm c} + f\times\kappa_\lambda^{\rm c, T}$ for the Sun,
and Figs. \ref{opcalpboodepth} and \ref{opcalpboowave} show the same for Arcturus. 
Because the near UV extinction of both objects is dominated by the same sources,
it does not seem likely that the missing $\kappa_\lambda$ source represented by
$f\times\kappa_\lambda^{\rm c, T}$ is directly related to any of the extinction
sources shown in Figs. \ref{opcsundepth} through \ref{opcalpboowave2}.  {\it Ie.}, 
it does not seem feasible that $f\times\kappa_\lambda^{\rm c, T}$ represents an
error in the calculation of any of the standard, well known $\kappa_\lambda^{\rm c}$
sources because, presumably, an error of the magnitude required to fit the 
near UV $f_\lambda$ level of Arcturus ($f\approx 1$) would also have a significant
impact on the Sun, whereas for the Sun $f\approx 0.15$.

\section{Conclusions}

 One unanticipated conclusion of our investigation is that there is a
surprisingly large amount of scatter ($\approx 10\%$ if one includes 
the \citet{necklabs} data set) in the measured
absolute $f_\lambda$ distribution for $\lambda<4000$\AA~ of the Sun. 
This undermines somewhat our ability to precisely test models of 
solar type stellar atmospheres, particularly with regard to the 
calculation of the near UV extinction, $\kappa_\lambda$.  
We conclude that a NLTE model based on the "small" high quality line list with 
standard extinction is consistent with the observations, and provides 
an especially good fit to the data of \citet{lockwood92} and \citet{burlov95}.
The fainter near UV band data of \citet{necklabs} can be better fit if
one adopts a modest enhancement,
$0.15\times\kappa_\lambda^{\rm c, T}$, in the $3000-4200$\AA~ region. 
We are unable to compute NLTE
models with the ``big'' line list, but we infer from a comparison
of the LTE-small and the NLTE synthetic spectra that such a model 
would provide a closer match to the observed $f_\lambda$ distribution of
\citet{necklabs},
thus obviating the need for additional continuum extinction that has
been proposed previously on the basis of comparison to that particular
data set. 
However, synthetic spectra computed with the ``big'' line list
have been shown to be more discrepant to with the observed spectrum
than those computed with the ``small'' list.  The continuum-rectified 
high resolution
solar spectrum is not sensitive to variations in the value of 
$\kappa_\lambda^{\rm c}$ of the magnitude necessary to reconcile
the absolute $f_\lambda$ distribution with observations.

Interestingly, a NLTE model with standard extinction and an {\it ad hoc} 
``semi-empirical'' $T_{\rm kin}(\log\tau_{\rm 5000})$ structure that
is $\approx 400$ K cooler at a $\log\tau_{\rm 5000}$ of $-6$, our NLTE-cool model, 
is also
provides a better fit to the observed near UV $f_\lambda$ distribution of
\citet{necklabs}, and
is in good agreement with a much more carefully derived semi-empirical model
$T_{\rm kin}(\log\tau_{\rm 5000})$ structure found by \citet{holmul}.  However,
the NLTE-cool model provide a worse fit to the well-observed visible band spectrum,
and we do not believe that adjustment of the $T_{\rm kin}(\log\tau_{\rm 5000})$ structure 
of a horizontally homogeneous model is the most compelling resolution of the
solar UV flux problem.

The situation is much more clear for the red giant, Arcturus.  All
models, computed with any line list, grossly under-estimate the
observed near UV band absolute $f_\lambda$ distribution in the 
near UV.  The
standard $\kappa_\lambda^{\rm c}$ must be approximately doubled
to match $f_\lambda$ in the $3000 - 4200$ \AA~ range.  At the same time,
the enhancement of $\kappa_\lambda^{\rm c}$ also brightens the cores
of strong spectra features and the wings of strongly damped
spectra lines, thereby improving the fit of the synthetic spectrum to
the observed continuum-rectified high resolution spectrum.  We conclude 
that there is
compelling evidence that there is still significant extinction 
missing from the models of early K star atmospheres, and that 
the missing extinction is probably due to a continuous process
rather than line absorption.

Given that the value of $\kappa_\lambda^{\rm c}$ in the near UV in both
the Sun and Arcturus is dominated by extinction sources that are 
readily calculated accurately, at least in LTE, namely, 
H$^{\rm -}$ $b-f$, and, to 
a lesser extent at extreme depths or heights, \ion{H}{1}
$b-f$, and Thomson scattering, it seems unlikely that the discrepancy 
between observed and computed near UV $f_\lambda$ levels is due
to an inaccuracy in the extinction calculation of known sources, especially 
in Arcturus where the discrepancy to be accounted for is so large.  
Although NLTE effects are significant at height for both \ion{H}{1}
$b-f$ and the total ``metal'' $b-f$ extinction, the effect of
NLTE upon the magnitude of $\kappa_\lambda^{\rm c}$ is opposite 
for these two sources, causing NLTE deviations in the total 
extinction to cancel out significantly.  Moreover, neither of these 
two sources dominates the near UV $\kappa_\lambda^{\rm c}$ level around
continuum optical depth unity.  Therefore, it seems unlikely that the near UV
$f_\lambda$ problem is due to inadequacies in the treatment of 
NLTE alone, particularly for Arcturus.  However, we note that the
computation of H$^{\rm -}$ $b-f$ extinction only accounts for 
NLTE effects indirectly in the contribution of some, but not all,
of the H and $e^{\rm -}$ sources in chemical equilibrium.

By contrast, LTE and NLTE models of Procyon both reproduce the UV 
band $f_\lambda$ distribution of Procyon down into the IUE LWR 
$\lambda$ range.  If
anything, there may be marginal evidence that models with standard
extinction predict slightly
too {\it little} flux in the near UV.  Admittedly, these three stars provide
a sparse set of data points for mapping out the dependence of any
inadequacy in standard extinction calculations on stellar parameters.
However, what is needed to improve the situation is a much larger
set of standard stars for which there is both 1) high quality 
absolute spectrophotometry, $f_\lambda(\lambda)$, down to 3500, 
if not 3000, \AA, and high quality high resolution (sufficient to
resolve spectral line profiles) continuum-rectified spectroscopy over 
a broad $\lambda$
range in the near UV, at least down to the limit for ground based astronomy 
($\approx 3600$ \AA).  
The sparseness of our stellar sample notwithstanding, comparison
of the {\it ad hoc} $\kappa_\lambda^{\rm c}$ enhancement factor, $f$, 
for Procyon, the Sun, and Arcturus leads us to conclude that, if
the near UV $f_\lambda$ discrepancy is due to an unaccounted for
continuum extinction source, then that source becomes increasingly
important with decreasing $T_{\rm eff}$ in the MK spectral class range 
from class mid-F to early K.  We note that \citet{short94} did compare
the measured $f_\lambda$ distribution of Arcturus with that
of two other K1.5-K2 III stars for which good quality data were available
(but for which high resolution near UV spectroscopy was {\it not} available),
and concluded that Arcturus is representative of stars of its
spectral class in the near UV band.

The dependence of $f$ on $T_{\rm eff}$ suggests a molecular origin, and
\citet{short94} proposed the molecular photo-dissociation 
of metal hydride diatomic molecules that are
relatively abundant in the atmospheres of late-type stars and that have
lower dissociation energies than more common molecules, often
corresponding to $\lambda$ values in the 3000 to 4000 \AA~ range.  
Unlike photo-ionization, photo-dissociation leads to a broad
$\kappa_\lambda$ feature without a sharp edge because the upper,  
pre-dissociation, electronic state has a complex structure. 
However, this hypothesis is difficult to test quantitatively 
because values of the cross sections for relevant metal hydride
molecules (MgH, SiH, FeH), either measured or calculated, are
not reported in the molecular physics literature.

\acknowledgments

CIS is grateful for NSERC Discovery Program grant 103815, a New Opportunities 
grant from CFI, and funding from NSRIT. 
The NSO/Kitt Peak FTS data used here were produced by NSF/NOAO.

\clearpage

\begin{deluxetable}{clllll}
\tablecaption{Stellar parameters adopted.\label{stelpar}}
\tablehead{\colhead{Object} & \colhead{$T_{\rm eff}$ (K)} & \colhead{$\log g$ (cm s$^{-2}$)} & \colhead{$[{{\rm A}\over{\rm H}}]$} & \colhead{$\xi_{\rm T}$ (km s$^{-1}$)} & \colhead{$l/H_{\rm P}$} }
\startdata
Arcturus & 4300 & 2.0    & -0.7 & 2.0 & 2.0 \\
Sun      & 5777 & 4.4377 & 0.0  & 1.0 & 1.0 \\
Procyon  & 6510 & 3.95   & 0.0  & 2.0 & 1.0 \\
\enddata
\end{deluxetable} 

\clearpage

\begin{deluxetable}{cc}
\tablecaption{Enhancement factor, {\it f}, of the {\it ad hoc} $\kappa_\lambda^{\rm c}$ enhancement used in the formula $\kappa_\lambda^{\rm c} + f\times\kappa_\lambda^{\rm c, T}$ (see text). \label{tabfudge}}
\tablehead{\colhead{Object} & \colhead{{\it f}} }
\startdata
Arcturus & 1.00 \\
Sun      & 0.15 \\
Procyon  & 0.00 \\
\enddata
\end{deluxetable}

\clearpage

\begin{deluxetable}{cccccc}
\tablecaption{Sun: RMS ($\sigma$) values for goodness of fit of synthetic
to observed $f_\lambda$ distributions. \label{tabsunstats}}
\tablehead{\colhead{} & \multicolumn{5}{c}{Data set} \\
\colhead{Model} & \colhead{Neckel\tablenotemark{a}} & \colhead{Burlov\tablenotemark{b}} & \colhead{Lockwood\tablenotemark{c}} & \colhead{Thuillier\tablenotemark{d}} & \colhead{All data}}
\startdata
LTE-big          & 0.048 & 0.083 & 0.082 & 0.062 & 0.071 \\ 
LTE-small        & 0.040 & 0.082 & 0.084 & 0.056 & 0.068 \\
LTE-small-extra  & 0.059 & 0.118 & 0.123 & 0.093 & 0.102 \\
NLTE             & 0.074 & 0.050 & 0.039 & 0.042 & 0.053 \\
NLTE-extra       & 0.027 & 0.074 & 0.076 & 0.046 & 0.060 \\
NLTE-cool        & 0.048 & 0.092 & 0.093 & 0.066 & 0.077 \\
\enddata
\tablenotetext{a}{\citet{necklabs}}
\tablenotetext{b}{\citet{burlov95}}
\tablenotetext{c}{\citet{lockwood92}}
\tablenotetext{d}{\citet{thuillier03}} 
\end{deluxetable}

\clearpage

\begin{deluxetable}{ccccc}
\tablecaption{Same as Table \ref{tabsunstats}, but for Arcturus. \label{tabarctstats}}
\tablehead{\colhead{} & \multicolumn{4}{c}{Data set} \\
\colhead{Model} & \colhead{Burnashev\tablenotemark{a}} & \colhead{Glushneva\tablenotemark{b}} & \colhead{Kharitonov\tablenotemark{c}} & \colhead{All data}}
\startdata
LTE-big         & 0.25 & 0.32 & 0.34 & 0.30 \\ 
LTE-small       & 0.31 & 0.38 & 0.41 & 0.37 \\
LTE-small-extra & 0.20 & 0.20 & 0.19 & 0.20 \\
NLTE            & 0.46 & 0.54 & 0.56 & 0.52 \\
NLTE-extra      & 0.12 & 0.15 & 0.14 & 0.14 \\
\enddata
\tablenotetext{a}{\citet{burn}}
\tablenotetext{b}{\citet{glushneva84}}
\tablenotetext{c}{\citet{kharitonov78}}
\end{deluxetable}

\clearpage



\clearpage

\clearpage

\clearpage

%

%

%

\clearpage

\clearpage

%
%
%
%

\clearpage

\clearpage



%
%
%

\clearpage

\clearpage

%

%

%

\clearpage

\clearpage

%

\clearpage

\clearpage

%

%

\clearpage

\clearpage

%

\clearpage

\clearpage

\end{document}